\documentclass[a4paper,12pt]{article}
\usepackage{graphicx}
\usepackage{bm}
\def\be{\begin{equation}}
\def\ee{\end{equation}}
\def\bea{\begin{eqnarray}}
\def\eea{\end{eqnarray}}
\usepackage{latexsym}
\usepackage{dcolumn}
\usepackage{epsfig,amssymb,euscript}
\usepackage{amsmath}
\usepackage{array,calc,epsfig}

\usepackage[noadjust]{cite}

\usepackage{chngcntr}

\usepackage[pdfpagelabels]{hyperref}

\usepackage{color}

\renewcommand{\to}{\rightarrow}
\renewcommand{\l}{\lambda}

\def\be{\begin{equation}}
\def\ee{\end{equation}}
\def\ba{\begin{eqnarray}}
\def\ea{\end{eqnarray}}
\def\nb{\nonumber}
\def\p{\partial}

\def\ff{\phi}

\def\d{\delta}

\def\l{\lambda}

\def\o{\omega}
\def\O{\Omega}

\def\q{\quad}

\def\mc{\mathcal}

\def\ra{\rightarrow}

\newcommand{\pr}[1]{\left(#1\right)}
\newcommand{\pq}[1]{\left[#1\right]}

\counterwithin*{equation}{section}

\begin{document}
\baselineskip=15.5pt
\pagestyle{plain}
\setcounter{page}{1}
\newfont{\namefont}{cmr10}
\newfont{\addfont}{cmti7 scaled 1440}
\newfont{\boldmathfont}{cmbx10}
\newfont{\headfontb}{cmbx10 scaled 1728}
\renewcommand{\theequation}{{\rm\thesection.\arabic{equation}}}
\renewcommand{\thefootnote}{\arabic{footnote}}

\vspace{1cm}
\begin{titlepage}
\vskip 2cm
\begin{center}
{\Large{\bf Bubble Wall Velocity at Strong Coupling}}
\end{center}

\vskip 10pt
\begin{center}
Francesco Bigazzi$^{a}$, Alessio Caddeo$^{b,a}$, Tommaso Canneti$^{c}$, Aldo L. Cotrone$^{a,c}$
\end{center}
\vskip 10pt
\begin{center}
\vspace{0.2cm}
\textit {$^a$ INFN, Sezione di Firenze; Via G. Sansone 1; I-50019 Sesto Fiorentino (Firenze), Italy.
}\\
\textit {$^b$ Galileo Galilei Institute for Theoretical Physics, INFN National Center for Advanced
Studies, Largo E. Fermi, 2, 50125 Firenze, Italy.
}\\
\textit{$^c$ Dipartimento di Fisica e Astronomia, Universit\'a di Firenze; Via G. Sansone 1;\\ I-50019 Sesto Fiorentino (Firenze), Italy.
}
\vskip 20pt
{\small{
bigazzi@fi.infn.it, caddeo@fi.infn.it, tommaso.canneti@stud.unifi.it, cotrone@fi.infn.it}
}

\end{center}

\vspace{25pt}

\begin{center}
 \textbf{Abstract}
\end{center}

\noindent 

Using the holographic correspondence as a tool, we determine the steady-state velocity of expanding vacuum bubbles nucleated within chiral finite temperature first-order phase transitions occurring in strongly coupled large $N$ QCD-like models. 
We provide general formulae for the friction force exerted by the plasma on the bubbles and for the steady-state velocity.
In the top-down holographic description, the phase transitions are related to changes in the embedding of $Dq$-${\bar Dq}$ flavor branes probing the black hole background sourced by a stack of $N$ $Dp$-branes. 
We first consider the Witten-Sakai-Sugimoto $D4$-$D8$-$\bar D8$ setup, compute the friction force and deduce the equilibrium velocity. 
Then we extend our analysis to more general setups and to different dimensions.
Finally, we briefly compare our results, obtained within a fully non-perturbative framework, to other estimates of the bubble velocity in the literature.

\end{titlepage}

\newpage
\tableofcontents
\section{Introduction}
Temperature driven cosmological first-order phase transitions are an exciting field of research in beyond the Standard Model physics. First-order phase transitions proceed through the nucleation of true vacuum bubbles within a metastable plasma. The bubbles expand due to the pressure gradient, collide and transfer energy to the surrounding plasma. Such violent inhomogeneous processes may be the source of stochastic gravitational wave (GW) backgrounds which could be within reach of present and near-future experiments \cite{Caprini:2019egz,Hindmarsh:2020hop}. 

The GW spectrum depends on two sets of parameters. The first set is composed of quantities which, in principle, can be determined from the underlying quantum field theory (QFT) model by means of static (Euclidean) computations. Examples are the bubble nucleation rate and the nucleation temperature, which can be computed from the on-shell effective action describing the ``bounce'', a Euclidean solution of the QFT equations of motion which interpolates between the true\footnote{In this context, the ``true'' vacuum is by construction the one inside the bubble, far away from the bubble wall, typically close to the true minimum of the potential.}  vacuum inside the bubble and the false vacuum outside. Further examples are thermodynamical quantities like the total energy released by the transition and the effective number of relativistic degrees of freedom.

The second set of parameters is formed by out-of-equilibrium quantities like the asymptotic bubble wall velocity and the efficiency factors measuring the kinetic energy transferred to the plasma. 
A precise computation of these quantities may reveal to be difficult even in perturbation theory.

If the phase transition happens within a strongly coupled QFT (as it is often conjectured to be the case for dark sectors) it may be hard to access both sets of parameters. Perturbative QFT tools and quasi-particle descriptions like Boltzmann kinetic theory 
have a limited regime of validity. Top-down holography has been recently employed to compute all the
Euclidean GW parameters in a class of QCD-like theories in  \cite{Bigazzi:2020phm, Bigazzi:2020avc} (partially inspired by the analysis in \cite{creminelli}). To the best
of our knowledge, no first principle computation of the out-of-equilibrium parameters in
strongly coupled systems is available in the literature. The aim of this work is to make progress in this direction.

We will focus on the steady-state bubble wall velocity. In principle, it can be computed by requiring that the total friction force exerted by the plasma on the expanding bubble is equilibrated by the pressure gradient. 
This topic has been the subject of an intense analysis in the literature. Results based on fluctuation-dissipation theorems \cite{Khlebnikov:1992bx,Arnold:1993wc} are complemented by perturbative computations in specific models based on kinetic theory \cite{Dine,Liu,Moore:1995ua, Moore:1995si,Espinosa:2010hh,Konstandin:2014zta,Bodeker:2017cim,Dorsch:2018pat,Mancha:2020fzw,Hoeche:2020rsg,Vanvlasselaer:2020niz,Cai:2020djd} as well as by the analysis of hydrodynamic effects \cite{Konstandin:2010dm,Balaji:2020yrx}.
Several regimes are envisaged depending on the bubble wall velocity and the friction force. If the velocity is smaller than the speed of sound of the plasma, the steady state is described by a deflagration. The opposite, supersonic, regime is called detonation. Hybrid regimes of supersonic deflagration happen to occur too. Finally, if the friction force is negligible, the bubble runs away reaching the speed of light.

In the present work, we will consider the friction force and the steady-state velocity for a class of strongly coupled QCD-like models having a dual top-down holographic description. 
The models are $SU(N)$ gauge theories with $N_f\ll N$ massless flavors\footnote{The masses of the flavors can be different from zero as long as they are smaller than the dynamically generated scale.} and describe the low-energy dynamics of a stack of $N$ $Dp$-branes and $N_f$ $Dq$-$\bar Dq$ ``flavor'' branes. The dual holographic description is provided by the near horizon solution sourced by the $Dp$-branes with $N_f$ flavor brane probes. The backreaction of the latter on the background can in fact be neglected in the $N_f\ll N$ limit where the flavors are quenched. The deconfined phase of the $SU(N)$ theory is described by a dual black hole background. The models we will consider feature a chiral first-order phase transition in the deconfined phase.\footnote{First-order transitions in the flavor sector of strongly coupled gauge theories in the deconfined phase are a quite general feature in top-down holography. Relevant examples not belonging to the class considered in the present work are for example \cite{Mateos:2006nu,Kobayashi:2006sb,Mateos:2007vn,Mateos:2007vc}. 
}
The latter can be seen as a two-component plasma, with a gluonic and a flavor part.

We will start our analysis from the QCD-like Witten-Sakai-Sugimoto (WSS) model \cite{Witten:1998zw, Sakai:2004cn}, featuring $N$ $D4$-branes wrapped on a circle and $N_f$ $D8$-$\bar D8$ branes. The bounce solution for the chiral phase transition has been studied in \cite{Bigazzi:2020phm} (where an analogous analysis has been performed for the confining-deconfining transition) and the results of the analysis have been used in \cite{Bigazzi:2020avc} to estimate the related GW spectra, using phenomenological relations for the out-of-equilibrium parameters. 

In the present work, we will consider an expanding bubble in the asymptotic steady state.
Rather than determining the complicated analytic form of the steady state, we consider a simplified configuration that should capture its main properties.
In this way, we will be able to provide an analytical estimate of the friction force exerted by the plasma and of the steady-state equation determining the wall velocity. By solving the equation, we will study the velocity as a function of the temperature. 
Finally, we will extend this analysis to more general $Dp$-$Dq$-$\bar Dq$ setups, determining the friction force and the steady-state condition fixing the wall velocity. 
In this paper we focus on the velocity of the bubble wall, leaving the study of the fluid dynamics for a future investigation.\footnote{Note that due to the $N_f \ll N$ condition, even though the bubble transfers energy to the plasma, we are completely neglecting its effects on the gluonic part of the plasma, which plays the role of a reservoir, unaltered by the dynamics of the process.}

Our analysis suggests that the steady-state condition can be written in a universal way in all the models we consider.  In particular, we will argue that for all the cases explored in the present paper, the total friction force (per unit area) due to the plasma, will be given (in $\hbar=c=K_B=1$ units) by
\be
\frac{F}{A} = C_d\, \frac{T_{boost}}{T_c}\, w_{f}(T_{boost})\, v + p_f(T_{boost})-p_f(T)\equiv \frac{F_d}{A}+p_f(T_{boost})-p_f(T)\,.
\label{introfa}
\ee
In the last step, we have defined the ``drag force'' per unit area $F_d/A$. The related drag coefficient $C_d$ is expressed in terms of the pressure $p_{glue}$ and the enthalpy density $w_{glue}$ (or equivalently in terms of the speed of sound $c_{s,glue}$) of the gluonic part of the plasma,\footnote{In the probe approximation it is not possible to distinguish the gluonic contribution to pressure and enthalpy from their total value. That is, in the approximation we are considering $p_{glue}/w_{glue}=p_{total}/w_{total}$.} as 
\be
C_d =  2\pi \frac{p_{glue}}{w_{glue}} \kappa_{c}=2\pi \frac{c_{s,glue}^2}{(1+c_{s,glue}^2)}\kappa_{c}\,,
\label{cdintro}
\ee
$\kappa_c$ being a numerical model-dependent coefficient, typically in the range $0.15-0.3$, related to the ratio between the critical temperature for the phase transition $T_c$ and the chiral symmetry breaking scale. In formula (\ref{introfa}), $w_f(T)$ and $p_f(T)$ are the enthalpy density and the pressure of the false vacuum and $v$ is the bubble wall velocity in the steady state. 
Moreover, $T_{boost}$ is a velocity-dependent boosted temperature, defined by
\be
T_{boost}= \gamma^{2/a}\,T\,,
\ee
where $\gamma=(1-v^2)^{-1/2}$ is the Lorentz factor and the coefficient $a$ gives the scaling with temperature $\rho\sim T^a$ of the energy density $\rho$ of the background gluonic plasma (for example, $a=6$ in the WSS model, where  $\rho \sim T^6/M_{KK}^2$, $M_{KK}$ being the dynamical scale.).

Furthermore, we will see that, for all the models explored in this paper, the steady-state bubble wall velocity, in the frame of the bubble center, will be deduced from the zero-force condition
\be
p_t(T)-p_f(T)\equiv \Delta p= \frac{F}{A}\,, 
\label{introfapfp}
\ee
where $p_t$ is the pressure of the true vacuum.
In other words, taking formula (\ref{introfa}) into account, the bubble wall velocity will be given by the solution of
\be
\boxed{
v = C_d^{-1} \frac{T_{c}}{T_{boost}} \frac{p_t (T) - p_f (T_{boost})}{w_f (T_{boost})}
}
\label{velocitygene}
\ee

This work is organized as follows. In section \ref{sec:WSS} we will review the main features of the WSS model and the analysis of the bounce solution for the WSS chiral transition. In section \ref{sec:WSSsteady} we will compute, within the WSS model, the drag force exerted by the plasma on the bubble in the steady state, obtain the zero-force condition determining the bubble wall velocity and study the latter as a function of the temperature. In section \ref{sec:general} we will extend our analysis to more general setups and deduce eq.~(\ref{cdintro}) for the drag coefficient. In section \ref{sec:compare} we briefly compare our findings with other proposals in the literature and draw our conclusions. Complementary material can be found in the appendices.
The computations of this paper are for the most part standard manipulations of probe branes in top-down holography; for the reader's convenience, we will summarize the main points at the beginning of each section.

Note added: when this work was in preparation, we became aware of the work \cite{BCN}, where the wall velocity is investigated in a (bottom-up) holographic model.\footnote{A class of Euclidean GW parameters in the same model has been recently studied in \cite{Ares:2020lbt}.} 
The numerical results in that paper point towards a linear relation for the velocity in the small velocity regime. Our equation (\ref{velocitygene}), which holds for generic values $v\leq 1$ of the wall speed, gives an analogous linear relation in the $v\ll1$ limit. 
 
\section{The WSS chiral transition}
\label{sec:WSS}

In this section we describe the Witten-Sakai-Sugimoto (WSS) model and its phase diagram focusing on chiral symmetry breaking \cite{Witten:1998zw,Sakai:2004cn,Aharony:2006da}.
In the deconfined phase, the model has a dual description as a black hole background (\ref{firstansatz}) encoding the properties of the gluonic vacuum.
The flavor degrees of freedom which will be the object of our investigation are dual to probe $D8$-branes embedded in such a background.

The branes have two branches for large values of the holographic radial direction, each branch supporting a $U(N_f)$ flavor symmetry.
The asymptotic distance $L$ of the two branches in formula (\ref{L1}) is the crucial parameter for the phase transition: being dual to a quartic interaction of the quarks \cite{Antonyan:2006vw}, it sets the (inverse) scale of chiral symmetry breaking.
In the dual brane perspective, there are two possible phases: in the first one, corresponding to preserved chiral symmetry and dominating at large temperatures, the two branches of the branes reach independently the horizon of the black hole (the ``disconnected'' configuration with action (\ref{sdisco}));
in the second phase, dominating at small temperatures, the two branches of the branes join smoothly before reaching the horizon (the ``connected'' configuration with action (\ref{sco})), reducing the chiral symmetry to the diagonal $U(N_f)$ subgroup and so corresponding to chiral symmetry breaking.   
The critical temperature for the phase transition can be determined by comparing the free energies in the two phases and turns out to be
\be
T_c \approx \frac{0.1538}{L}\,.
\ee
 
When the temperature is lowered through $T_c$, bubbles of true vacuum, corresponding to connected brane configurations, are nucleated in the false vacuum, corresponding to the disconnected configuration.
The bubbles interpolating between the two phases at the nucleation time have been studied in \cite{Bigazzi:2020phm}.
An example of such solutions is reported in figure \ref{fig1}.

In the rest of this section, we provide details on the above description. 
 
\subsection{The model}

The WSS model, at low energy, is a $3+1$ dimensional non-supersymmetric $SU(N)$ gauge theory coupled to $N_f$ fundamental flavors and a tower of adjoint Kaluza-Klein massive fields. The gauge theory describes the infrared dynamics of $N\gg1$ $D4$-branes wrapped on a circle of radius $R_4 = M_{KK}^{-1}$ along a compact space direction $x_4$. Fundamental matter fields correspond to two stacks of  $N_f$ ``flavor'' $D8$-anti-$D8$-branes ($D8-\bar D8$) placed at different points on the circle. The holographic description of the model simplifies in the quenched approximation $N_f\ll N$ and in the strong coupling limit $\lambda\gg1$, where $\lambda$ can be seen as the 't Hooft coupling at the compactification scale $M_{KK}$. In this regime, the theory is holographically mapped into the near-horizon classical gravity background sourced by the $D4$-branes, with the flavor branes acting as probes. 

The WSS model features two kinds of first-order transitions at finite temperature. One is a confinement-deconfinement transition, corresponding to a Hawking-Page transition between a ``solitonic'' background at low temperature and a black hole background at high temperature. The critical temperature for such a transition is $T_{c,conf} = M_{KK}/2\pi$. In the confined phase, chiral symmetry is always spontaneously broken, and this fact is nicely accounted for by the joining of the two asymptotically separated stacks of $D8-\bar D8$ branes at a certain value $u=u_J$ of the holographic radial coordinate. We will refer to the related $U$-shaped $D8$-brane profile as the \emph{connected configuration}. 

In the deconfined phase at $T>M_{KK}/2\pi$, depending on the distance $L$ between the $D8$ and the $\bar D8$ branes, another first-order transition can occur. In particular, if $L M_{KK}\ge 0.966$, chiral symmetry is always restored and the two stacks of branes remain \emph{disconnected} extending all along the radial direction $u$ down to the black hole horizon. If  $L M_{KK} < 0.966$, instead, chiral symmetry is broken if $T< T_c$ and it is restored if $T>T_c$, where $T_c \sim 0.1538 L^{-1}$. 
The latter (and analogous ones in more general $Dp$-$Dq$-$\bar Dq$ setups) is the chiral phase transition we will focus on in this work.

\subsection{The two phases}
\label{sub2}
The deconfined phase of the WSS model is holographically described by a black hole background with metric
\be
\label{firstansatz}
ds^2 = \pr{\frac{u}{R}}^{3/2} \pq{- f_T(u) dt^2 + dx_i dx_i +  d x_4 ^2 } +\pr{\frac{R}{u}}^{3/2} \pq{\frac{du^2}{f_T(u)} +u^2 
d \O _4 ^2} \ ,
\ee
where
\be
f_T(u) =1- \frac{u_{T}^3}{u^3}\,.
\ee
Here $t, x_i, i=1,2,3$ are the non-compact Minkowski directions where the gauge theory lives, $x_4\simeq x_4 +2\pi M_{KK}^{-1}$ is the compactified direction, $u$ is the radial coordinate holographically mapped into the renormalization group energy scale and the remaining directions fill a compact four-sphere $S^4$. The metric has an event horizon at $u=u_T$. The background also supports a running dilaton and a four-form Ramond-Ramond field strength given by
\be
e^\ff=g_s \pr{\frac{u}{R}}^{3/4}\ ,\quad F_4= \frac{3 R^3 }{g_s} \o_4\ , \quad R^3=\pi g_s N l_s ^3\ ,
\ee
where $\omega_4$ is the volume form of the $S^4$ sphere and $g_s, l_s$ are the string coupling and the string length. The parameter $u_T$ is related to the temperature $T$ by
\be
u_T = \frac{16 \pi ^2}{9} R^3 T ^2 \ .
\label{utth}
\ee
The map between string parameters and field theory ones is completed by
\be
g_s l_s = \frac{1}{4 \pi} \frac{\l}{M_{KK }N } \ , \q \q \q  \frac{R^3}{l_s ^2} = \frac{1}{4} \frac{\l}{M_{KK}}\ , 
\label{holomaps}
\ee
where $\lambda$ is the 't Hooft coupling at the scale $M_{KK}$.

The free energy $\cal F$ of the Yang-Mills plasma dual to the above background can be analytically computed by means of the standard holographic relation ${\cal F}=T S_E$, where $S_E$ is the (renormalized) on-shell Euclidean ten-dimensional action. From this the whole thermodynamic observables can be deduced. For instance, the energy density reads
\be
\rho_{glue}=5\frac{2^6 \pi^4 }{3^7} \l N^2  \frac{T^6}{M_{KK}^2}\,.
\label{rhoradglue}
\ee   
Note that in the deconfined phase the model exhibits the thermodynamics of a six-dimensional theory since compact dimensions effectively open up. We will consider theories that have different dimensionalities in the deconfined phase in section \ref{sec:general}.
We will use the Minkowski signature as in (\ref{firstansatz}) to explore real-time physical quantities, like the steady states we will discuss in the following. When focusing on equilibrium physics and in the computation of bubble nucleation rates, the Euclidean continuation of the background will be considered, as usual.  

The $D8$-flavor branes extend along the Minkowski directions, the four-sphere and $u$, with a profile $x_4 = x_4(u)$. The Euclidean Dirac-Born-Infeld action for each brane is
\begin{equation}
S_{DBI}=\frac{T_8}{g_s}\int d^9x \left(\frac{u}{R}\right)^{-3/2}u^4 \sqrt{1 + f_T(u) \left(\frac{u}{R}\right)^3 (\partial_u x_4)^2}  \ ,
\label{SDBI1}
\end{equation}
where $T_8 = (2\pi)^{-8} l_s^{-9}$ is the brane tension. The Euler-Lagrange equation for the embedding reads 
\begin{equation}
\partial_u\left(\frac{u^4 f_T(u) \left(\frac{u}{R}\right)^{3/2} (\partial_u x_4)}{\sqrt{1 + f_T(u) \left(\frac{u}{R}\right)^3 (\partial_u x_4)^2} }\right) = 0 \ ,
 \label{simple1}
\end{equation}
and it has to be solved imposing the boundary condition $x_4(u\rightarrow\infty)\rightarrow\pm L/2$. This reflects the asymptotic setup with $D8$ and $\bar D8$ brane stacks separated by a distance $L$ along the compact circle. The simplest solution of (\ref{simple1}), $x_4=\pm L/2$, describes the disconnected straight brane-antibrane pair entering the black hole horizon. This is the setup corresponding to the phase where the classical chiral symmetry $U(N_f)\times U(N_f)$, realized by the gauge symmetry on the flavor branes (global symmetry in the dual QFT), is unbroken.

The phase with broken chiral symmetry corresponds to non-trivial U-shaped solutions where the brane and antibrane join at some radial position $u=u_J>u_T$ where $x_4'(u_J)=\infty$.
For this case, we can solve (\ref{simple1}) as
\begin{equation}
\frac{u^4 \sqrt{f_T(u)}}{\sqrt{1 + \left(f_T(u) \left(\frac{u}{R}\right)^3 (\partial_u x_4)^2\right)^{-1}}} = u_J^4 \sqrt{f_T(u_J)} \ .
\label{simple2}
\end{equation}
If we rescale the coordinates as
\begin{equation}
x_4 = x \,  u_T^{-1/2} R^{3/2} = x \frac{3}{4\pi T} \ , \qquad u = y\,u_T   \ ,  \qquad u_J = y_J\,u_T\,,
\label{redef}
\end{equation}
such that
\begin{equation}
f_T(u) \equiv f_T = 1 - y^{-3}  \ , \qquad  f_T(u_J)  \equiv f_{TJ} = 1 - y_J^{-3} \ ,
\label{redef2}
\end{equation}
we can rewrite equation (\ref{simple2}) as
\begin{equation}
\partial_y x = \left[f_T  y^3 \left(\frac{y^8f_T}{y_J^8 f_{TJ}} - 1 \right)  \right]^{-1/2} \ .
\label{yofx1}
\end{equation}
The distance between the brane and the antibrane along $x_4$ can thus be computed as
\begin{equation}
L = \int_{worldvolume} dx_4 = 2 \int_{u_J}^\infty \frac{dx_4}{du} du = 2 \frac{3}{4\pi T} \int_{y_J}^\infty
 \left[f_T  y^3 \left(\frac{y^8f_T}{y_J^8 f_{TJ}} - 1 \right)  \right]^{-1/2} dy \ ,
 \label{L1}
\end{equation}
where the factor of 2 arises from adding up both sides of the ``U". The parameter $L$ is related to the chiral symmetry breaking scale in the dual QFT. In the limit $L\ll \pi M_{KK}$, for instance, the latter is given by
\be
\label{fchil}
f_{\chi}^2 \approx 0.1534 \frac{\lambda N}{32 \pi^3} \frac{1}{M_{KK} L^3}\ .
\ee
In order to understand in which cases the U-shaped profile is energetically preferred to the disconnected configuration, we have to holographically compute the related free energy difference $\Delta {\cal F}=\Delta S_{DBI} T$, where $S_{DBI}$ is the Euclidean on-shell DBI action.  Let us consider $N_f$ $D8-\bar D8$ branes and define
\be
K\equiv \frac{T_8}{g_s}N_f\frac{V_{3}}{T}V_{S^4} R^{3/2} u_T^{7/2}=V_3\frac{8\pi^2}{3^8}\lambda^3N N_f\frac{T^6}{M_{KK}^3}\,,
\ee
where $V_3$ is the (infinite) volume of 3d spatial directions and $V_{S^4}=8\pi^2/3$ is the volume of the internal four-sphere.
The on-shell DBI action for the disconnected configuration is
\begin{equation}
 S_{DBI}|_{d} = 2 K \int_1^{y_{cut}} y^{5/2} dy  \ ,
\label{sdisco}
\end{equation}
where $y_{cut}$ is a UV cutoff.
For the connected on-shell configuration, using (\ref{yofx1}) we get
\begin{equation}\label{sco}
S_{DBI}|_c = 2K \int_{y_J}^{y_{cut}} y^{5/2} \left(1-\frac{y_J^8 f_{TJ}}{y^8 f_T} \right)^{-1/2} dy \ .
\end{equation}
The difference $\Delta S_{DBI} =S_{DBI}|_c - S_{DBI}|_d $ is not divergent and the UV cut-off can be
safely removed. It reads
\begin{equation}
\frac{\Delta S_{DBI}}{K}\equiv \Delta\tilde S= 2 \int_{y_J}^{\infty} y^{5/2} \left[\left(1-\frac{y_J^8 f_{TJ}}{y^8 f_T} \right)^{-1/2} - 1\right] dy
-  \frac47 (y_J^{7/2} -1) \ .
\label{eqDS}
\end{equation}
A simple numerical analysis of the above expression shows that $\Delta S_{DBI}>0$ for $y_J<y_{\chi SB} \approx 1.3592$. In this case the disconnected configuration is preferred and chiral symmetry is preserved. 
Conversely,  $\Delta S_{DBI}<0$ for $y_J>y_{\chi SB}$ and the connected configuration is energetically favored.
 The value of $y_{\chi SB}$ corresponds to  $(L T)_{c}\approx 0.1538$. At $T=T_c$ a first-order transition occurs between the two phases. 
 
Let us stress that the validity of the probe approximation requires (see e.g.~\cite{Bigazzi:2014qsa})
\be 
\epsilon_{f,T} \equiv \frac{\l^2 N_f T} {6 \pi^2 N M_{KK}} \ll 1\,.
\ee
In this limit holography precisely provides the thermodynamic observables in the flavor sector. The pressure difference between the broken and the symmetric phase is given by
\be
\Delta p = - \frac{T}{V_3}\Delta S_{DBI} = - \frac{8\pi^2}{3^8}\lambda^3N N_f\frac{T^7}{M_{KK}^3}\Delta \tilde S\,.
\label{pressgrad}
\ee
In the chirally symmetric phase (which in the following will be termed the ``false'' vacuum) the energy density is given by
\be
\label{energyfalse}
\rho_{f} =\frac{2^6 \pi^2}{7\cdot 3^7} \lambda^3 N_f N \frac{T^7}{M_{KK}^3}\,,
\ee
while that of the broken phase (the ``true'' vacuum in the following) can be deduced from
\be
\rho_{t} = \rho_{f} - (1 - T \partial_T )\Delta p\,.
\ee
\subsection{Flavor brane bubbles}

If we consider a cosmological evolution where we start at $T>T_c$ and then decrease the temperature until we cross the critical temperature $T_c$, spherical bubbles of the broken phase (the true vacuum) form within the symmetric plasma (the false vacuum). Due to the isometries of the background, the related bounce should be $O(3)$-symmetric and corresponds to a space-dependent solution $x_4 = x_4 (u, \rho)$ of the Euclidean DBI Euler-Lagrange equations, which interpolates between the true vacuum at the center of the bubble at $\rho=\sqrt{x^i x^i}=0$ and the false vacuum at $\rho\rightarrow\infty$. An approximate solution was numerically found in \cite{Bigazzi:2020phm} using a variational approach. 

\begin{figure}[htb]
\begin{center}
\includegraphics[width=0.495\textwidth]{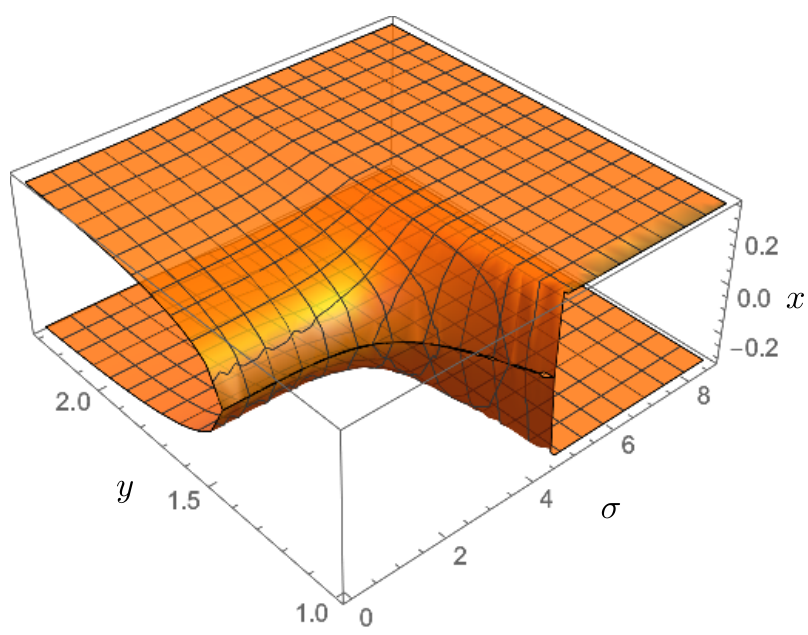}
\end{center}
\caption{Plot of the thin-wall bounce profile $x(y,\sigma)$ at the nucleation time, for $\tilde L = 0.62$ where $\tilde L\equiv 4\pi L T/3$ and $(x,\sigma)=(4\pi T/3)(x_4, \rho)$. The configuration smoothly interpolates between U-shaped profiles at $\sigma=0$ and disconnected branes at $\sigma \to \infty$.}
\label{fig1}
\end{figure}

A picture of the typical profile of a thin-wall solution, emerging when the nucleation temperature is close to the critical one, is given in figure \ref{fig1}. Here a U-shaped configuration very close to the true vacuum exists for a finite range of $\rho$ which then rapidly evolves into the false vacuum.
Let us stress the fact that the main analysis of this paper is expected to be valid also for thick-wall nucleated bubbles, i.e. for bubbles whose thickness is comparable to their radius \emph{at nucleation time $t_n$}. In fact, the steady state we will consider is a late-time configuration ($t \gg t_n$) where the radius of the bubble has grown very large, while there is no reason a-priori for the thickness to grow at the same rate. 

Using the on-shell bounce action, the nucleation rate and the corresponding nucleation and percolation temperatures have been computed in \cite{Bigazzi:2020phm, Bigazzi:2020avc}. 

\section{Steady-state bubbles in WSS}
\label{sec:WSSsteady}

In this section, we consider the asymptotic late-time physics of the bubbles. 
Once sufficiently large bubbles are nucleated, they start expanding due to the pressure gradient between the interior and the exterior regions. 
It is a crucial issue to understand whether there is a runaway expansion, driving the bubble to the maximal allowed velocity $v=1$, or if, due to the friction force exerted by the plasma, the bubble reaches a steady state with a different velocity. 

In the context of the WSS model, we set up the calculation of the friction force, starting from what is commonly referred to as the ``drag force'' in holography, and ultimately of the bubble velocity, obtained from the balancing of the friction force with the pressure gradient.
The idea is that in its motion the nucleated bubble, whose dual brane description is depicted in figure \ref{fig1}, reaches a steady state after a transient period of acceleration.
In this evolution, the profile of the brane dual to the bubble gets distorted by the forces acting on it: the pressure difference and the friction of the plasma, which is encoded in the transfer of momentum from the brane to the black hole horizon.
Thus, we envisage a profile of the bubble steady state whose cartoon is reported in figure \ref{trailing3d}.
The part of the brane which moves and gets in contact with the horizon, ``trailing'' behind the bubble wall (the bottom part in the picture on the left of figure \ref{trailing3d}), is the relevant one for the calculation of the drag force.

The equation of motion (\ref{eom}) for the steady state is a highly non-trivial, non-linear PDE, whose solution is beyond the scope of this paper. Here our goal is to derive analytic estimates for the wall velocity, so we will mainly focus on a simplified ``\emph{rectangular}'' version of the steady state, as described below. 
Of course, this entails the crucial assumption that the rectangular configuration correctly captures the main physical properties of the full solution,
``and the reader should be warned that if it is false, all my [our] conclusions are garbage'' \cite{Coleman:1977py}.   
By considering this rectangular configuration, which has trivial shape in the transverse direction $x_4$, we will be able to compute the drag force by a standard holographic procedure, with the result reported in (\ref{fdragexpr}).

On the other hand, the pressure difference between the interior and exterior of the bubble is encoded in the tension contributions of the different parts of the brane.
These can be calculated explicitly
and their balance with the friction force gives formula (\ref{effesuahere}), which in the Introduction we have written as (\ref{introfapfp}) and finally as (\ref{velocitygene}), and whose dependence on the temperature is reported in the plot in figure \ref{figvelnew}.  

Let us begin by introducing the ansatz for the trailing wall.

\subsection{The trailing wall}
\label{subsecsteadystateeq}

Let us consider an asymptotic state at time $t \gg t_n$, where $t_n$ is the nucleation time.
The radius of the bubble has grown accordingly to be very large.
Assuming that the thickness of the bubble wall has not grown, or that it has grown at a smaller rate (there is no a-priori reason for the thickness to grow at the same rate), we end up in a ``thin-wall'' situation, where the radius is much larger than the thickness, irrespectively of the initial situation at nucleation time.
Moreover, at relativistic velocities the Lorentz contraction of the bubble thickness becomes relevant.
Accordingly, we are going to employ the ``thin-wall approximation'', where the thickness is considered to be zero and so the transition from the inside and the outside of the bubble is abrupt.

Moreover, since the radius is very large, the curvature of the bubble is very small.
In such a case, we can approximate the bubble wall profile in the quantum field theory with a plane (say, along $(x_1, x_2)$) moving in the orthogonal direction, say $x_3\equiv z$. In this case, taking $x_4=x_4(t,z,u)$, the action takes the form
\be\label{dbisteady}
S_{DBI}= -\frac{T_8}{g_s} \int d^9x \left(\frac{u}{R}\right)^{-3/2}u^4 \sqrt{1 + f_T(u) \left(\frac{u}{R}\right)^3 (\partial_u x_4)^2 + (\partial_z x_4)^2- f_T(u)^{-1}(\partial_t x_4)^2}  \,.
\ee 
The steady-state ansatz is taken to be of the form
\be
x_4 (t, z, u) = x_4(z-v t, u)\,,
\ee
so that $\partial_{t} x_4 = -v\, \partial_{z} x_4$.  Actually, since we will be interested in determining the momentum flow in the direction of motion, it will be more convenient to describe the wall by means of the inverse embedding function
\be
z = v t + \xi(u,x_4)\,,
\label{embe}
\ee 
such that $z_w = v t$ will correspond to the position of the bubble wall in the dual field theory.

Some intuition on the bubble wall profile in the thin-wall limit at $t=t_n$ is provided by figure \ref{fig1}. Between the connected solution on the left and the disconnected one on the right, there is a sharp separation surface orthogonal to the bubble radius. This can be seen as a $D8$-brane piece extended along $u$ and $x_4$ and orthogonal to $z$ (the equivalent of $\sigma$ in the figure), which separates the connected from the disconnected branch. 

The action related to the ansatz (\ref{embe}) is given by\footnote{We use the notation $\partial_4 \equiv \partial_{x_4}$.}
\be
S = -\frac{k}{L} \int dt\,du\,dx_4 \left(\frac{u}{R}\right)^{-3/2}u^4 \sqrt{1 + (\partial_4\xi)^2+ f_T(u) \left(\frac{u}{R}\right)^3 (\partial_u \xi)^2 - f_T(u)^{-1}v^2}\,,
\label{sdbinewtot}
\ee
where
\be
k \equiv \frac{T_8}{g_s} A\,L\,V(S^4) \,, \quad A\equiv \int dx_1 dx_2\,.
\label{kdef}
\ee
The Euler-Lagrange equations are
\be \label{eom}
\partial_u \pi^u_{\xi} + \partial_4 \pi^4_{\xi} = 0\,,
\ee
where 
\be
\pi_{\xi}^u =k\frac{u^4 f_T(u) \left(\frac{u}{R}\right)^{3/2} \partial_u \xi}{\sqrt{1 + (\partial_4\xi)^2+ f_T(u) \left(\frac{u}{R}\right)^3 (\partial_u \xi)^2 - f_T(u)^{-1}v^2}}\,,
\label{pixinew}
\ee
and
\be
\pi_{\xi}^4 = k\frac{u^4 \left(\frac{u}{R}\right)^{-3/2} \partial_4 \xi}{\sqrt{1 + (\partial_4\xi)^2+ f_T(u) \left(\frac{u}{R}\right)^3 (\partial_u \xi)^2 - f_T(u)^{-1}v^2}}\,,
\label{pi4new}
\ee
and the overall constant is chosen for future convenience. In the steady state the total momentum along $z$ is conserved. This zero-force condition is derived in section \ref{app:subb} by integrating equation (\ref{eom}) over $u$ and $x_4$, and imposing boundary conditions that make the wall connect to the connected and the disconnected pieces of the full brane configuration. Before doing that, in the next section, we study the trailing wall by itself within what we will call the ``rectangular approximation''.

\subsection{The drag force}  
\label{secdrag}

Taking inspiration from the thin-wall picture of the bubble at $t=t_n$, let us approximate (only) a part of the  trailing wall with a rectangular $D8$-brane extended along $x_4$ and along $u$. This surface starts from $u_J$ (i.e.~the tip of the connected configuration) at a fixed position $z=z_w$ and asymptotically approaches the horizon $u_T$.

Moreover, let us assume that the embedding profile $\xi(u,x_4)$ is trivial along $x_4$,  such that $\pi^4 _\xi =0$.  
This is what we call ``rectangular approximation''. In this section, we study the trailing wall as a separate entity, without taking into account the fact that in the complete steady-state configuration, it attaches to the connected and the disconnected branches. In sections \ref{seccompsteady} and \ref{subsec:wsszeroforce}, we will compute the total friction force taking into account the presence of these other branches. 

The ansatz describing the steady-state motion of this rectangular wall will thus be
\be
z = v\, t + \xi(u)\,.
\label{wallans}
\ee

Using the ansatz (\ref{wallans}), the DBI action (\ref{sdbinewtot}) reduces to
\be
S _w= -\frac{k}{L} \int dt\,du\,dx_4 \left(\frac{u}{R}\right)^{-3/2}u^4 \sqrt{1 + f_T(u) \left(\frac{u}{R}\right)^3 (\partial_u \xi)^2 - f_T(u)^{-1}v^2}\,,
\label{sdbinew}
\ee 
where $\xi' = d\xi/du$. Now the idea is to compute the drag force exerted by the plasma on the wall in close analogy with the holographic computation of the drag force on a heavy quark moving in a quark-gluon plasma \cite{Herzog:2006gh,Gubser:2006bz}. In that case, the problem reduced to solving for the steady-state profile of a trailing string on a black hole background. A similar analysis of the steady-state motion of defects in AdS and more general $Dp$-brane backgrounds can be found in \cite{Janiszewski:2011ue, FuiniJohnF:2011aa}. 

\begin{figure}[htb]
\begin{center}
\includegraphics[width=0.495\textwidth]{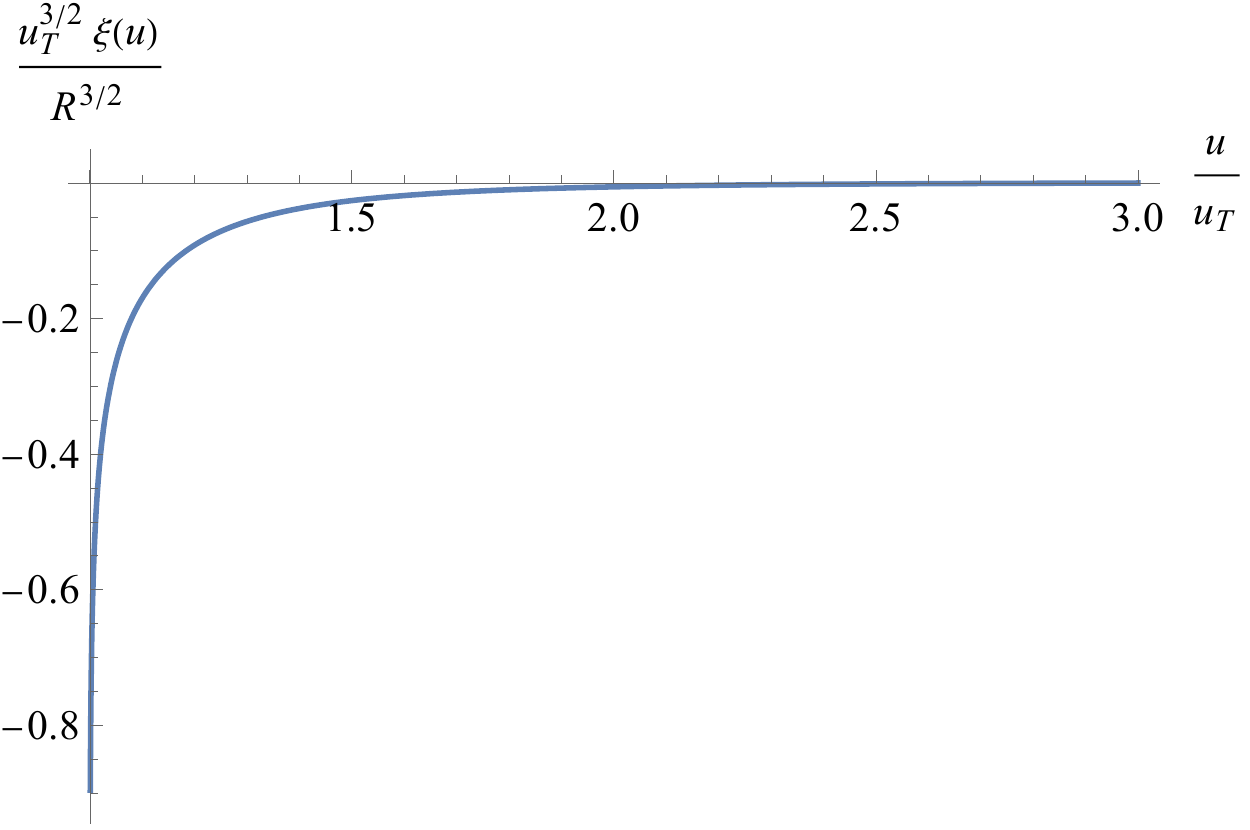}
\end{center}
\caption{The trailing solution of eq.~(\ref{xiprime}) for $v=0.5$ and $u_{J}=3u_T$. We have imposed $\xi(u_J)=0$. Notice that $\xi'(u_J)\neq0$ despite what one could guess from the picture.}
\label{figprof}
\end{figure}

Let us first notice that from the action above it follows that $\pi_{\xi}^u$ is constant and it is given by
\be
\pi_{\xi}^u = k \frac{u^4 f_T(u) \left(\frac{u}{R}\right)^{3/2} \xi'}{\sqrt{1 + f_T(u) \left(\frac{u}{R}\right)^3 (\xi ')^2 - f_T(u)^{-1}v^2}}\,.
\label{pixi}
\ee
Hence,
\be
\xi' = \pi_{\xi}^u\left(\frac{u}{R}\right)^{-3/2}f_T(u)^{-1/2}\sqrt{\frac{1-f_T(u)^{-1}v^2}{k^2 f_T(u) u^8 - (\pi_{\xi}^u)^2}}\,,
\label{xiprime}
\ee
where we have fixed the sign ambiguity in order for the above expression to be consistent with (\ref{pixi}).
In formula (\ref{xiprime}) both the numerator and the denominator under the square root change sign at 
\be
u_c(v) = \frac{u_T}{(1-v^2)^{1/3}}\,,
\label{ucv}
\ee
where
\be
\pi_{\xi}^u = \pi_{\xi}^u(u_c)= k\, u_T^4\frac{v}{(1-v^2)^{4/3}}\,.
\label{pifin}
\ee 
In the above expression, which sets the value of the constant momentum, the overall positive sign is selected by requiring the solution of (\ref{xiprime}) to trail towards the horizon behind the moving wall. In such a way the wall momentum flows towards the horizon as required by the classical picture of the black hole background.  
A picture of the typical resulting profile is given in figure \ref{figprof}.

Notice that the simplest ansatz $z=vt$, which trivially solves the equations of motion, is not a consistent solution for all the values of $u$: indeed, in this case the square root in (\ref{sdbinew}) would reduce to $\sqrt{1 - f_T(u)^{-1}v^2}$, which becomes imaginary for every $v \neq 0$ if $u<u_c(v)$. The fact that the wall embedding crosses the critical value $u_c(v)$ is precisely what implies the existence of a non-zero drag force.

The radial position $u_c(v)$ has a very precise geometrical meaning: it is the radial position of the horizon of the induced metric on the $D8$-brane wall. The relevance of this induced horizon for the physics of the trailing strings has been largely commented on in the literature.

Notice that since $u_T \propto T^2$ the relation (\ref{ucv}) between $u_c(v)$ and $u_T$ can also be read as a relation
\be
T_{boost} = \frac{T}{(1-v^2)^{1/6}}\,,
\label{Tboost}
\ee
defining the boosted temperature in the WSS model. 

Using the result (\ref{pifin}) and following the same reasoning as in \cite{Gubser:2006bz} we can thus deduce the drag force.
Since $\pi^4_{\xi}=0$, the world-volume current of
spacetime energy-momentum carried by the brane is given just by $\pi^u_{\xi}$, which is constant according to (\ref{eom}).
Hence, when calculating the momentum flow $dp_{z}/dt$ which goes down the brane and is transferred to the horizon, we have to integrate $\pi^u_{\xi}$ over the relevant time interval, evaluating it at any desired value of $u$.
Accordingly, the force just reads
\be
F_d= \frac{dp_{z}}{dt} = \pi_{\xi}^u = \frac{T_8}{g_s}A\, L\, V(S^4) u_T^4\frac{v}{(1-v^2)^{4/3}}\,.
\ee
Rewriting this in terms of gauge theory quantities (and reinserting an overall $N_f$ factor) we get that the drag force per unit surface reads
\be\label{effesua}
\frac{F_d}{A}= \frac{2^5}{3^9} \pi^3 \lambda^3 N N_f  (L T) \frac{T^7}{M_{KK}^3}\frac{v}{(1-v^2)^{4/3}}\,.
\ee
Taking into account the expression (\ref{energyfalse}) for the energy density $\rho_f$ of the disconnected configuration i.e.~of the false vacuum, we conclude that the drag force per unit surface can be written as
\be
\frac{F_d}{A} = \frac{\pi}{3} L\, T_{boost}\, w_{f}(T_{boost})\, v \equiv C_d \,\frac{T_{boost}}{T_c}\, w_{f}(T_{boost})\, v\,,
\label{fdragexpr}
\ee
where $C_d\approx 0.16$ is a model-dependent drag coefficient, given by formula (\ref{cdintro}) (see section \ref{generalDpDqdrag} for a detailed interpretation of the latter), $T_c=0.1538 L^{-1}$ is the critical temperature for the phase transition and $w_{f}(T_{boost})$ is the enthalpy density ($w= \rho+ p$) of the false vacuum at the boosted temperature $T_{boost}$. 

As already remarked, in order to derive these results we have considered a simplified rectangular configuration, where $\partial_{4}\xi=0$ and so $\pi^4_{\xi}=0$.
Let us stress again that this is essentially a technical assumption, due to the extreme complexity of the problem.
As we will discuss in the following, at least the leading behavior of the true solution can be shown to be really independent on $x_4$ close to the horizon, where the friction is supposed to be localized.
Thus, what we are assuming is that the same behavior far from the horizon is sufficient to capture the main physics of the true solution.
It is possible that a rigorous solution of equation (\ref{eom}) would give corrections to the drag force calculated above.
Nevertheless, the neatness of formula (\ref{fdragexpr}) is such that we would be surprised if it would not be close to  the full result.

\subsection{The complete steady-state configuration}
\label{seccompsteady}

\begin{figure}[htb]
\begin{center}
\includegraphics[width=0.55\textwidth]{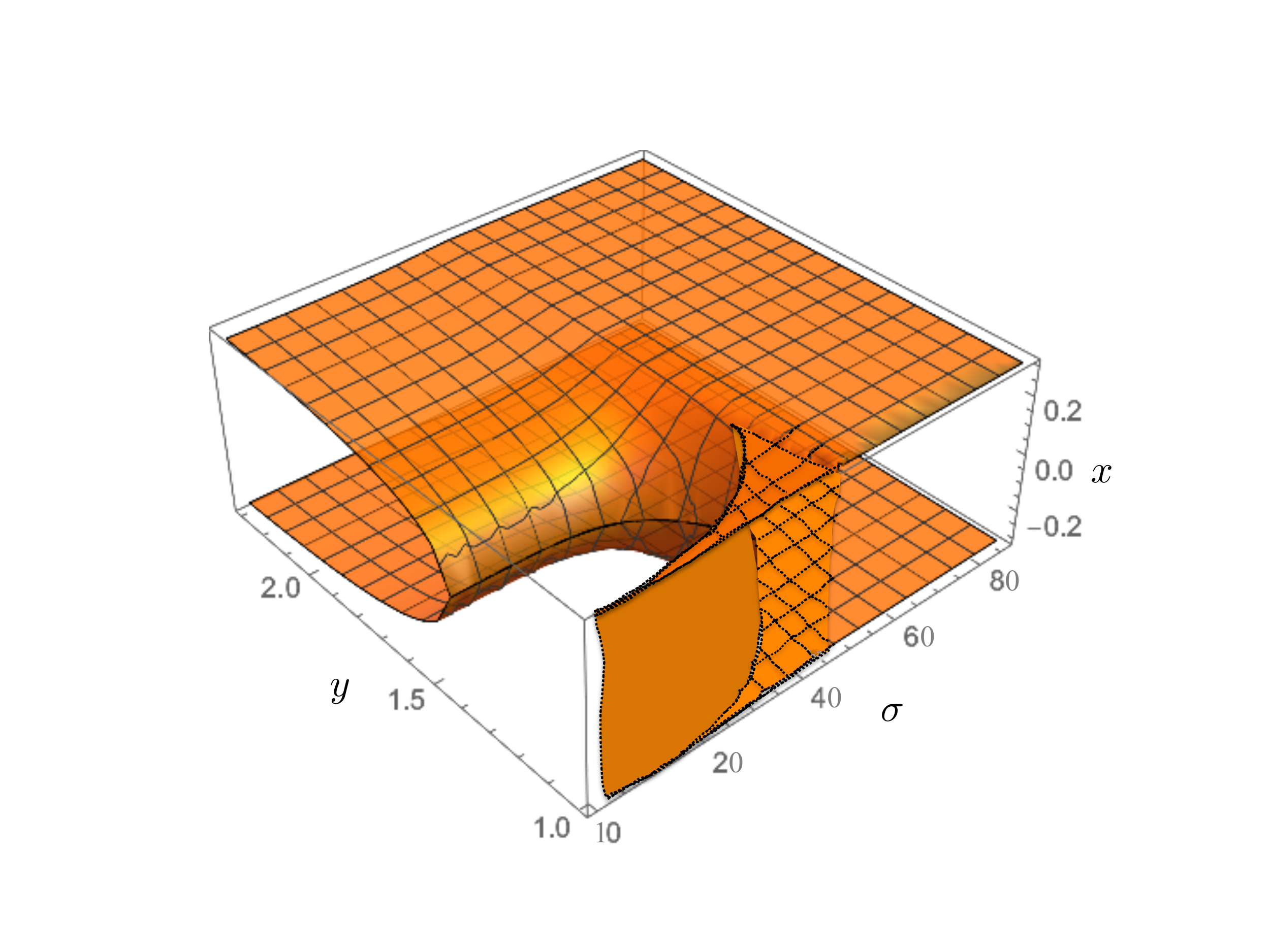}
\includegraphics[width=0.25\textwidth]{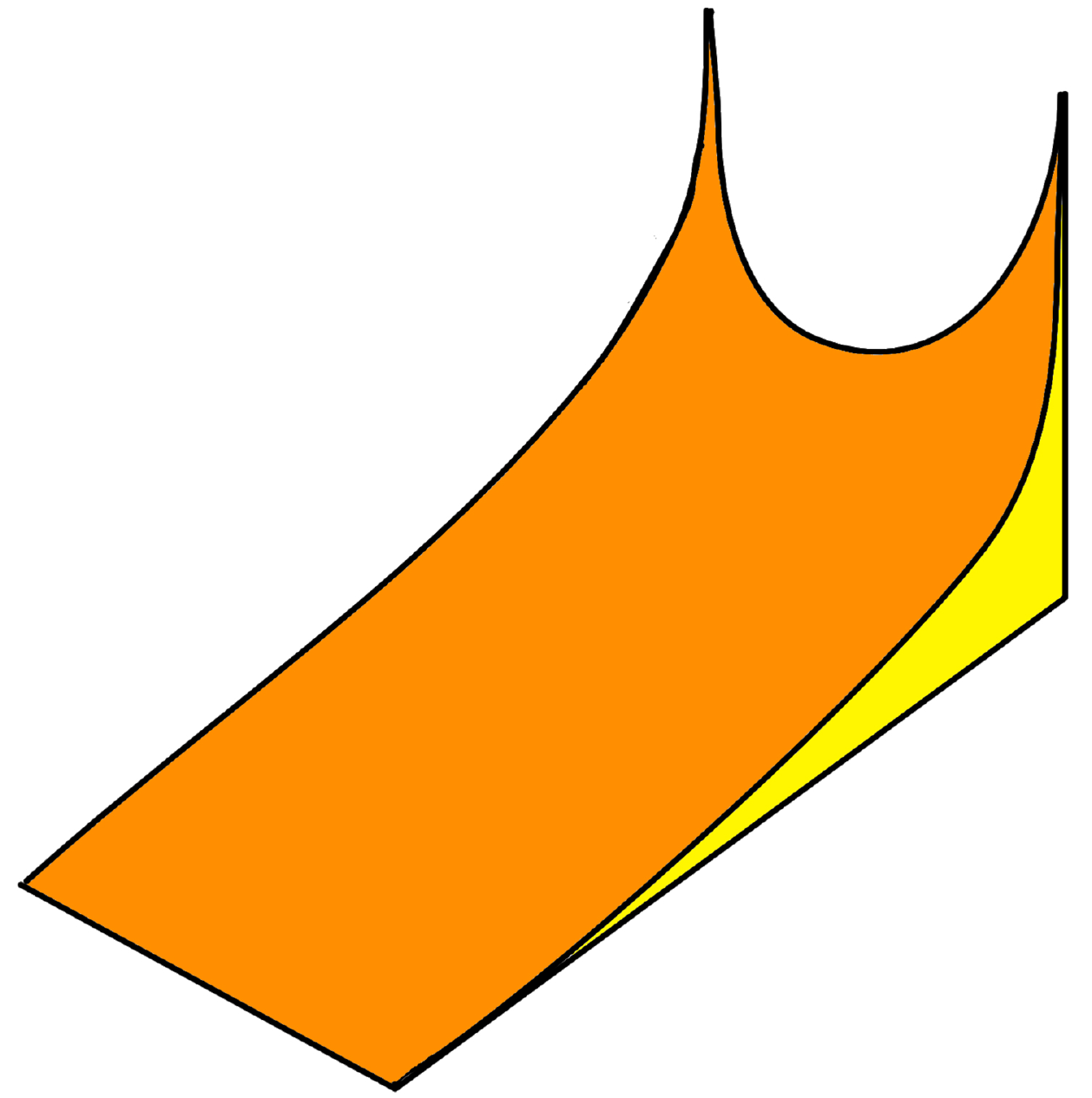}
\end{center}
\caption{On the left, a cartoon of a possible steady-state solution, where $y=u/u_T$ and $(x,\sigma)=(4\pi T/3)(x_4, z)$. On the right a very schematic rotated zoom on the trailing wall and its boundaries.}
\label{trailing3d}
\end{figure}

The simple surface we have focused on using the reduced ansatz $z = vt + \xi(u)$ has lost the important information of being part of the whole configuration which separates the true and the false vacua. This information can be recovered by considering the complete setup described by the action (\ref{sdbinewtot}). A cartoon of a possible complete steady-state solution 
is given in figure \ref{trailing3d}. 
In this picture, the trailing wall considered above is expected to arise as a backward continuation of the connected configuration. In the whole steady-state solution, the tip of the connected configuration at fixed time describes a curve that starts from $u_J$ at the bubble center ($z\rightarrow -\infty$). Then increasing $z$ the tip position decreases up to a point $u=u_*$ where $\xi'(u_*)=0$ and finally turns around asymptotically approaching the horizon position $u_T$ again at $z\rightarrow-\infty$, where, $\xi'(u_T)\rightarrow\infty$.  

The trailing wall in the picture has also two boundary slides (in yellow in the zoom in figure \ref{trailing3d}) at $x_4=\pm L/2$ where $\partial_4 z\rightarrow\pm\infty$. Here the $D8$-brane embedding coincides with that of the disconnected configuration. 
We expect the full solution of equation (\ref{eom}) to be a smoothed-out version of figure \ref{trailing3d}, where the sharp edges at $x_4 = \pm L/2$ are replaced by smooth surfaces.

The picture suggests that the trailing wall will be subject to two opposite forces: one is the friction force exerted by the plasma due to the trailing towards the horizon; the other emerges as a combination of two effects: the force due to the yellow slides, which, in order to minimize their area, will tend to move the wall towards the right; the force due to the connected configuration on top of the wall, which, in order to minimize its energy, will tend to move towards the horizon, hence driving the trailing wall towards the left. What we expect is that these two effects combine in a force (per unit surface) directed towards the right and given by the pressure gradient $\Delta p$ between the true vacuum on the left and the false vacuum on the right. Steady-state motion will then imply that $A\,\Delta p = F$. The aim of the following subsections is to show that these expectations are explicitly realized.

Before going on, let us try to justify one crucial assumption we are doing in drawing the pictures in figure \ref{trailing3d}, i.e.~that the intersection of the trailing wall with the horizon $u=u_T$ is always described by a rectangular curve made of a straight line along $x_4$ at $\xi\rightarrow-\infty$ and two lines along $\xi$ at $x_4=\pm L/2$. 

Let us consider the near-horizon behavior of the e.o.m.~(\ref{eom}), where the only assumption is that there exists a steady-state solution (i.e.~here we do not limit ourselves to the simplified rectangular configuration).
The momenta $\pi^u_\xi $ and $\pi^4 _\xi$ display the denominator
\be\label{squareroot}
\sqrt{1 + (\partial_4\xi)^2+ f_T(u) \left(\frac{u}{R}\right)^3 (\partial_u \xi)^2 - f_T(u)^{-1}v^2} \,.
\ee
Since at the horizon $f_T(u)$ vanishes, the $x_4$-independent negative term in $v^2$ blows up and the square root becomes imaginary.
In order to avoid this, the other two terms under the square root in (\ref{squareroot}) must compensate the diverging negative factor, so there are the two possibilities:
\begin{itemize}
\item $\partial_4 \xi \rightarrow \infty$ as $u \rightarrow u_T$. This means that for these points the intersection of the brane with the horizon is straight along the $z$ coordinate at fixed $x_4$. This happens for example at large $z$ where the embedding coincides with the disconnected configuration, where $x_4=\pm L/2$.
\item $\partial_u \xi \rightarrow \infty$ as $u \rightarrow u_T$. This means that for these points the intersection of the brane with the horizon is straight along the $x_4$ coordinate at fixed $z$.
\end{itemize} 

Let us consider the second possibility, since the first one is simple and solves automatically the e.o.m.~close to the horizon.
We do not have complete control of the solution close to the horizon, but from inspection of (\ref{squareroot}), one expects $\xi$ to display at least a  $x_4$-independent logarithmic divergence. 
We have verified that the equation of motion forbids other kind of divergences such as poles and square-root cuts.
As a result, the near-horizon limit of the true configuration $\xi(u,x_4)$ is expected to take the form
\bea\label{ansatz}
\xi(u,x_4) \sim \frac{R^{3/2} v}{3 u_T^{1/2}} \log (u-u_T)
+ f_0(x_4) + f_1(x_4) (u-u_T) + f_2(x_4) (u-u_T)^2 +....
\eea
This ansatz solves the e.o.m.~close to the horizon, with $f_2(x_4)$ expressed in terms of $f_0(x_4)$, $f_1(x_4)$ without further constraints. 
The form (\ref{ansatz}) means that the general solution of (\ref{eom}) goes to $z \rightarrow -\infty $ as it approaches the horizon, that is, the piece of the embedding close to the horizon and which is not straight along $z$ is far away at infinity: the configuration is trailing behind the wall.
Thus, all the steady-state solutions have an intersection with the horizon which is ``straight'': two straight lines along $z$ at fixed $x_4$ and a straight line along $x_4$ ``at $z \rightarrow -\infty$''.


\subsection{The zero-force condition}
\label{app:subb}
Let us consider the complete steady-state configuration represented in figure \ref{trailing3d}. In this section, we study how to embed the trailing wall configuration discussed in section \ref{subsecsteadystateeq} into the complete brane setup. In this way, we will be able to derive the zero-force condition that equates the total friction force to the pressure gradient. 

We take the thin-wall limit in which the separation between the connected and the disconnected parts is sharp. In this way, we can identify four brane branches, as also shown in figure \ref{trailing3d}. 
The first one (in orange in the right picture in figure \ref{trailing3d}) is the trailing wall. The second one is the connected branch, extending from $z \ra -\infty$ to the place where the trailing wall ends. Such connected piece has the U-shaped connected profile $u_{ss}(x_4)$, solution of equation (\ref{yofx1}). The third branch is the disconnected one, extending along the $z$ direction from the place where the trailing wall ends towards infinity. Finally, the brane configuration features two vertical slides (in yellow in figure \ref{trailing3d}) at $x_4 = \pm L/2$.

It is worth stressing that the domain of the trailing profile $z=z(u,x_4)$ is the region $D_C$ spanned by $x_4 \in [-L/2,L/2]$ and $u \in [u_T,C]$, where $C$ is the curve $u=u_{ss} (x_4)$.

According to the description above, we split the action as
\be
S = S_{conn} + S_{disc} + S_w + S_{sl}  \ ,
\ee
where
\begin{subequations}
\ba
S_{conn} &=&  -\frac{k}{L} \int dt \int dx_4 \int_{u_T}^{+\infty} du\, \delta(u-u_{ss}) \int_{-\infty}^{z_w(C)}dz \,{\cal L}_{c}\ , \\
S_{disc} &=&  -\frac{2 k}{L} \int dt \int dx_4 \, \delta(x_4-L/2) \int_{u_T}^{+\infty} du  \int ^{\infty} _{z_w(C)} dz \, \,{\cal L}_{d}\ ,\\
S_{sl}& =& - \frac{2k}{L}\int dt \int dx_4\, \delta(x_4-L/2) \int_{u_T}^{+ \infty} du  \int_{z_w(u,L/2)}^{z_w(C)} dz{\cal L}_d  \ , \\
\label{Sslides2b}
S_{w} &=& - \frac{k}{L}\int dt \int dx_4 \int_{u_T}^{+\infty}  du \, \Theta (u_{ss}-u)  {\cal L}_w(\partial z_w) \ .
\ea
\end{subequations}
Here, $\mc{L}_c$ and $\mc{L}_d$ denote the Lagrangian densities for the connected and the disconnected configurations, which can be read from the actions (\ref{sco}), (\ref{sdisco}), whereas $\mc{L}_w$ is the trailing wall Lagrangian density associated to the action (\ref{sdbinewtot}).

Let us take the variation of the action with respect to $z_w (u,x_4)$, 
\ba
\label{variationtotalactionnonrectangular}
\d S &=& \frac{1}{L} \int dt \int dx_4 \int_{u_T}^{+\infty}  du\,  \Theta (u_{ss}-u) (\p_u \pi_\xi ^u + \p_4 \pi_\xi ^4)\, \d z_w (u,x_4) \nb \\
&-& \frac{1}{L} \int dt \int dx_4 \int_{u_T}^{+\infty} du\, \delta(u-u_{ss}) \pr{k \mc{L}_c + \pi_\xi ^u} \d z_w (C) \nb \\
&+&\frac{1}{L} \int d t \int dx_4 \pi_\xi ^u |_{u=u_T} \d z_w (u_T,x_4) \nb \\
&-& \frac{1}{L} \int d t \int_{u_T} ^{+\infty} du \pq{\pi_\xi ^4 (x_4,u) - k \mc{L} _d}_{x_4=L/2} \d z_w (u,L/2) \nb \\
&+& \frac{1}{L} \int d t \int_{u_T} ^{+\infty} du \pq{\pi_\xi ^4 (x_4,u) + k  \mc{L} _d}_{x_4=-L/2} \d z_w (u,-L/2)  \ ,
\ea
from which we read the Euler-Lagrange equation (valid on $D_C$)
\be
\label{ELquationalessioIIb}
\p_u \pi_\xi ^u + \p_4 \pi_\xi ^4  =0  \ .
\ee
We should also pay attention to the boundary terms. We do not impose Dirichlet boundary conditions along $x_4$ since the profile ends on the yellow slides of figure \ref{trailing3d} and since we also take the variation of $S_{sl}$. An analogous comment holds for the boundary condition on $C$. As a result, from the last two lines of (\ref{variationtotalactionnonrectangular}), we read
\be
\label{momentum4atboundary4b}
\pi ^4 _\xi (x_4=L/2,u) = - \pi ^4 _\xi (x_4=-L/2,u) = k\,\mc{L} _d \equiv k R^{3/2} u^{5/2}\ ,
\ee
while from the second line of (\ref{variationtotalactionnonrectangular}) we read
\be
\int dx_4 \int_{u_T}^{+\infty} du\, \delta(u-u_{ss}) \pi_\xi ^u (x_4,u) = - k \int dx_4 \int_{u_T}^{+\infty} du\, \delta(u-u_{ss}) \mc{L}_c \ .
\label{bdruj}
\ee
Integrating the Euler-Lagrange equation (\ref{ELquationalessioIIb}) over the whole domain $D_C$, we find
\be
\label{g1nonrectangular}
\frac{1}{L}\int dx_4\, \pi_{\xi}^{u}(u_T, x_4) = \frac{2k}{L}\int_{u_T}^{+\infty}du{\cal L}_d - \frac{k}{L} \int dx_4 \int_{u_T}^{+\infty} du\, \delta(u-u_{ss}) \mc{L}_c \ .
\ee
The right-hand side of eq.~(\ref{g1nonrectangular}) is proportional to the difference between the static U-shaped connected and disconnected on-shell Euclidean actions, i.e.~formula (\ref{eqDS}). In particular
\be
T\Delta S_E = - \frac{2k}{A L} \int d^3 x \int_{u_T}^{+\infty}du{\cal L}_d + \frac{k}{A L} \int d^3 x \int dx_4 \int_{u_T}^{+\infty} du\, \delta(u-u_{ss}) \mc{L}_c \ ,
\ee
where, as above, $A=\int d^2x$. Moreover, using the holographic relation between the above expression and the free energy in the dual QFT, we get
\be
T \Delta S_E = \Delta {\cal F} = \int d^3x \Delta f = - \int d^3 x \Delta p\,,
\ee
where $f$ is the free energy density and $\Delta p$ is the pressure difference between the true and false vacua.
Thus, equation (\ref{g1nonrectangular}) reads
\be
F\equiv\overline{\pi_{\xi}^u(u_T)} \equiv \frac{1}{L}\int dx_4\, \pi_{\xi}^{u}(u_T, x_4)  =  A\, \Delta p\,.
\label{expenonrect}
\ee
The left-hand side of equation (\ref{expenonrect}), i.e.~the (average along $x_4$) of the momentum flow towards the horizon, is interpreted as the \emph{total friction force} that the plasma exerts on the wall. Thus, equation (\ref{expenonrect}) relates the friction force to the pressure difference, and therefore gives the zero-force condition for the steady state.
As we will see in the next section, the total friction force can be written as the sum of the drag force computed in section \ref{secdrag} plus another contribution. 

\subsection{The bubble wall velocity}
\label{subsec:wsszeroforce}

Once the zero-force condition equation (\ref{expenonrect}) is derived, one needs to evaluate its left-hand side, namely the total friction force, 
since its right-hand side is known from equation (\ref{pressgrad}). In section \ref{secdrag}, we computed the drag force for the case in which we neglect that the trailing wall is part of the complete brane configuration. The aim of this section is to compute the total friction force for the case in which we take into account the complete brane setup. We will work in the rectangular approximation within which $\pi^4_\xi =0$, so that the trailing wall is rigid along the $x_4$ direction. Moreover we will approximate the connected part of the complete steady-state configuration with two vertical lines at $x_4 = \pm L/2$ and a horizontal line at $u=u_J$.
Details on the zero-force condition (obtained following the steps of section \ref{app:subb}) in this  approximation where also the connected portion of the brane inside the bubble is rectangular can be found in appendix \ref{explicitcheck2}. 

Within the above approximation, from the last two lines of (\ref{variationtotalactionnonrectangular}), we see that a source term for the equation of motion remains. In principle, the latter is a source localized at $x_4 = \pm L/2$. The rectangular approximation consists in taking this source as independent of $x_4$ since the tension of the brane along $x_4$ becomes infinite. As a result, the equation of motion $\p_u \pi^u _\xi=0$ valid in section \ref{secdrag}, where the trailing wall was studied by itself, gets deformed into
\be
\partial_u\pi_{\xi}^u = - 2\frac{k}{L} R^{3/2} u^{5/2}\,,
\label{eqpid}
\ee
once one takes into account that the trailing wall attaches to the connected and the disconnected pieces of the complete steady-state configuration. The term on the right is proportional to the Lagrangian density ${\cal L}_d$ of the disconnected configuration (\ref{sdisco}). 

The equation of motion is supplemented with the boundary condition (\ref{bdruj}), which in the rectangular approximation reduces to
\be
\label{bconditionatujrectangular}
\pi_\xi ^u (u_J) = - k u_J^4 \sqrt{f_T(u_J)}\equiv - k \mc{L}_{ch}(u_J)\,,
\ee
where one can recognize a term proportional to the Lagrangian density of the horizontal part of the  connected configuration in the rectangular approximation, trivially extended along $x_4$ at $u=u_J$ (this can be seen formally taking the $x_4'(u_J)\rightarrow\infty$ limit in eq.~(\ref{SDBI1})).
After integration of (\ref{eqpid}), we get
\be
\pi_{\xi}^u = -\frac47 \frac{k}{L} R^{3/2} ( u^{7/2} - u_*^{7/2})\ .
\label{pinte}
\ee
The position $u_*$ corresponds to the maximum value of $z$ reached by the connected part of the brane in figure \ref{trailing3d}. 
Using the boundary condition (\ref{bconditionatujrectangular}), from (\ref{pinte}) we get 
\be \label{qui0}
u_*^{7/2} = u_J^{7/2} - \frac{7L}{4 R^{3/2}}{\cal L}_{c h}(u_J) \ .
\ee
Note that $u_* \leq u_J$ if ${\cal L}_{c h}(u_J)>0$.
Equation (\ref{qui0}), in turn, can be rewritten as
\be \label{ustar}
u_*^{7/2} = u_T^{7/2} + \frac{7L}{4k R^{3/2}} A\, \Delta p \,,
\ee
so that it is apparent that $u_* \geq u_T$. Here $\Delta p$ is the pressure difference between the connected and disconnected phases in the approximated setup we are considering. 
It is proportional to the difference between the static rectangular connected and disconnected on-shell Euclidean actions. Actually, the vertical parts of the connected solution (extending between $u_J$ and $\infty$) cancel with the corresponding parts of the disconnected solutions, and, as a result
\be
k {\cal L}_{c\,h}(u_J) - \frac{2k}{L}\int_{u_T}^{u_J}du{\cal L}_d= - A \Delta p\,.
\ee
As the analysis of the previous subsection shows, in the more realistic case of a thin-wall setup with an actual U-shaped connected part of the configuration inside the bubble (instead of the rectangular one) with tip in $u_J$, the analysis will automatically replace the pressure gradient of the oversimplified example with the actual pressure gradient $\Delta p$ between the true and false vacuum as given by (\ref{pressgrad}). From now on, let us thus extrapolate our analysis to the more realistic setup and make use of formula (\ref{pressgrad}) when writing $\Delta p$. 
 
 We are in a position to derive the formula for the steady-state velocity. Plugging (\ref{eqpid}) into (\ref{xiprime}), we find
\be
\xi' = -\frac47 \frac{k R^{3/2}}{L}\left( u^{7/2} - u_*^{7/2}\right)\left(\frac{u}{R}\right)^{-3/2}f_T(u)^{-1/2}\sqrt{\frac{1-f_T(u)^{-1}v^2}{k^2 f_T(u) u^8 - \frac47 \frac{k R^{3/2}}{L}\left( u^{7/2} - u_*^{7/2} \right)^2}}\,,
\label{xiprimenew}
\ee 
so that $\xi'(u_*)=0$. At the radial position $u_c$ where the numerator and the denominator in the square root of (\ref{xiprimenew}) are zero, one has
\be
k f_T^{1/2} u_c^4 = - \frac47 \frac{k R^{3/2}}{L} (u_c^{7/2} - u_*^{7/2}) \,.
\ee
Explicitating this relation making use of (\ref{ustar}), (\ref{ucv}), we get (cfr.~(\ref{effesua})) the zero-force condition
\be\label{effesuahere}
\Delta p= \frac{2^5}{3^9} \pi^3 \lambda^3 N N_f  (L T) \frac{T^7}{M_{KK}^3} \left[ \frac{v}{(1-v^2)^{4/3}}  -\frac47 \frac{3}{4\pi L T}\left( 1- \frac{1}{(1-v^2)^{7/6}} \right)\right]\,,
\ee
or, using (\ref{pressgrad}),
\be\label{velhere}
-\frac{\Delta \tilde S}{\tilde L}= \left[ \frac{v}{(1-v^2)^{4/3}}  -\frac47 \frac{1}{\tilde L}\left( 1- \frac{1}{(1-v^2)^{7/6}} \right)\right]\,,
\ee
where
\be
\tilde L \equiv \frac{4\pi}{3}LT \eqsim 0.644\,\frac{T}{T_c}\,. 
\ee
The no-force condition (\ref{effesuahere}), determining the steady-state velocity, can equivalently be rewritten as
\be
\Delta p = \frac{F}{A} \equiv \frac{F_d}{A} + p_{f} (T_{boost}) - p_{f} (T)\,,
\label{zeroff}
\ee
or, using $\Delta p = p_t(T)-p_f(T)$, as
\be
\frac{F_d}{A}= p_t(T) - p_f(T_{boost})\,,
\label{fawssf}
\ee
where $F_d/A$ is the drag force per unit area as determined in the previous section in eq.~(\ref{fdragexpr}), $T_{boost}$ is the boosted temperature defined in (\ref{Tboost}) and $t,f$ stand for true and false vacua respectively. It is tempting to interpret the drag force as a purely out-of-equilibrium contribution to the total friction. The remaining piece in eq.~(\ref{zeroff}) could instead be seen as a local equilibrium term, possibly related to a heating of the plasma in front of the bubble wall, along the lines of what suggested in \cite{Konstandin:2010dm,Balaji:2020yrx}. We cannot exclude however that this apparent distinction is just an artefact of the quenched approximation we are adopting in this paper.
 
\begin{figure}[t!]
\begin{center}
\includegraphics[width=0.6\textwidth]{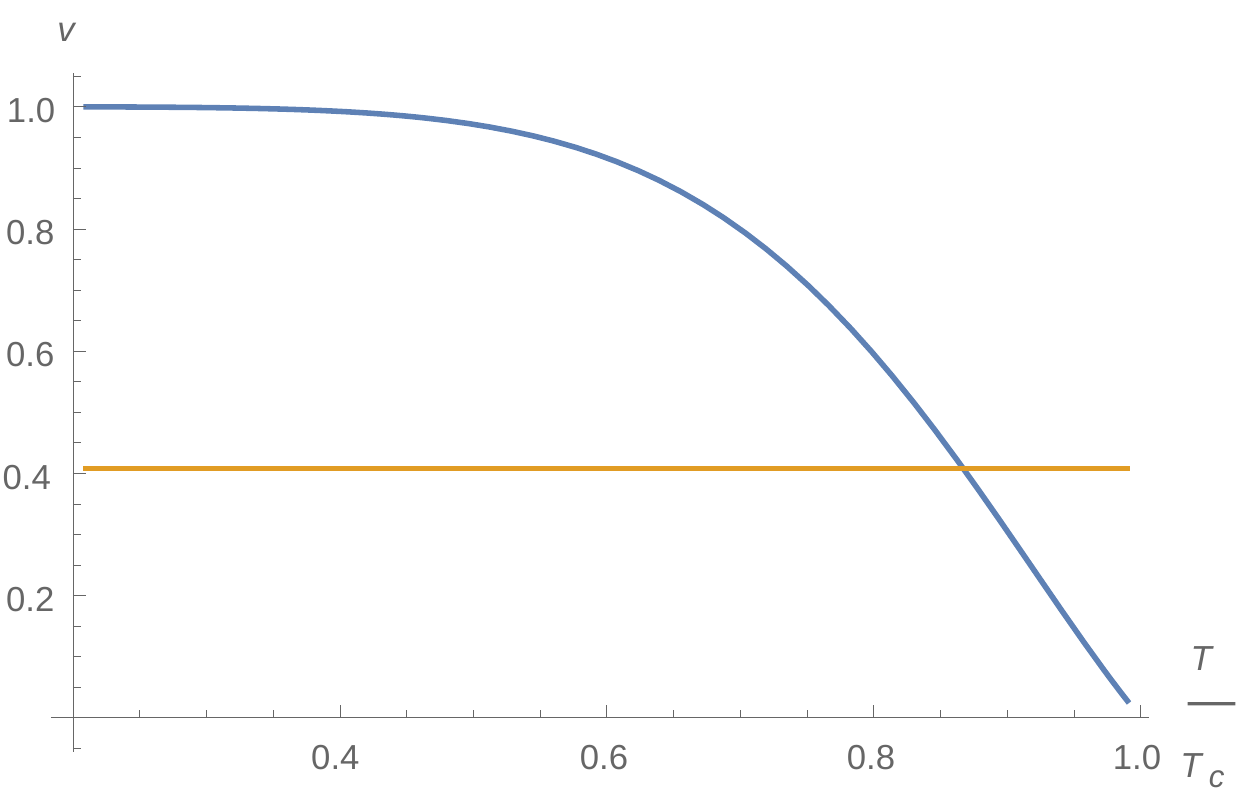}
\end{center}
\caption{The bubble wall velocity as a function of the relative temperature $T/T_c$ from formula (\ref{velhere}). The constant line is the speed of sound of the flavor sector in the false vacuum, $c_s^2=dp_f/d\rho_f=1/6$. 
}
\label{figvelnew}
\end{figure}

Using the above relations and formula (\ref{fdragexpr}) we get the implicit relation determining the wall velocity
\be
v = C_d^{-1} \frac{T_c}{T_{boost}} \frac{p_{t} (T) - p_{f} (T_{boost})}{w_{f}(T_{boost})}\,.
\ee
In figure \ref{figvelnew} one can see the behavior of the velocity as a function of the relative temperature $T/T_c$ from (\ref{velhere}).

Let us now look at the solution of (\ref{xiprimenew}). 
Let us introduce the dimensionless quantities $y \equiv u/u_T$ and
\be
\chi \equiv \frac{\xi}{L}\,,
\ee
so that the equation reads (here $'=\partial_y$)
\be
\chi' = -\frac{1}{\tilde L}\frac{y^{7/2} - 1 +\frac74 \Delta\tilde S}{y^{3/2}f_T(y)^{1/2}}\sqrt{\frac{1-f_T(y)^{-1}v^2}{\frac{7^2}{2^4}{\tilde L}^2 f_T(y) y^8 - \left(y^{7/2} - 1 +\frac74 \Delta\tilde S \right)^2}}\,.
\label{chiprimenew}
\ee

\begin{figure}[htb]
\begin{center}
\includegraphics[width=0.45\textwidth]{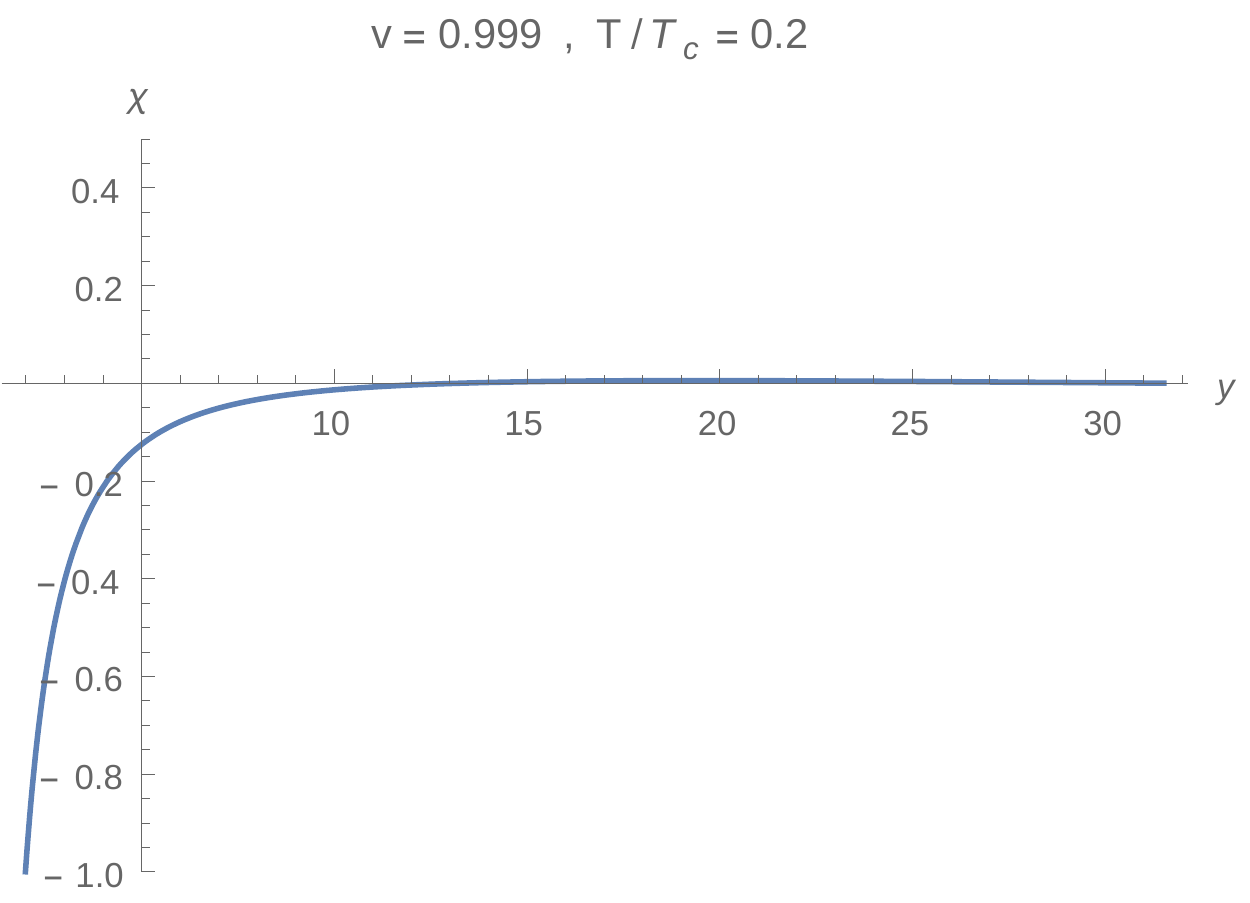}
\includegraphics[width=0.45\textwidth]{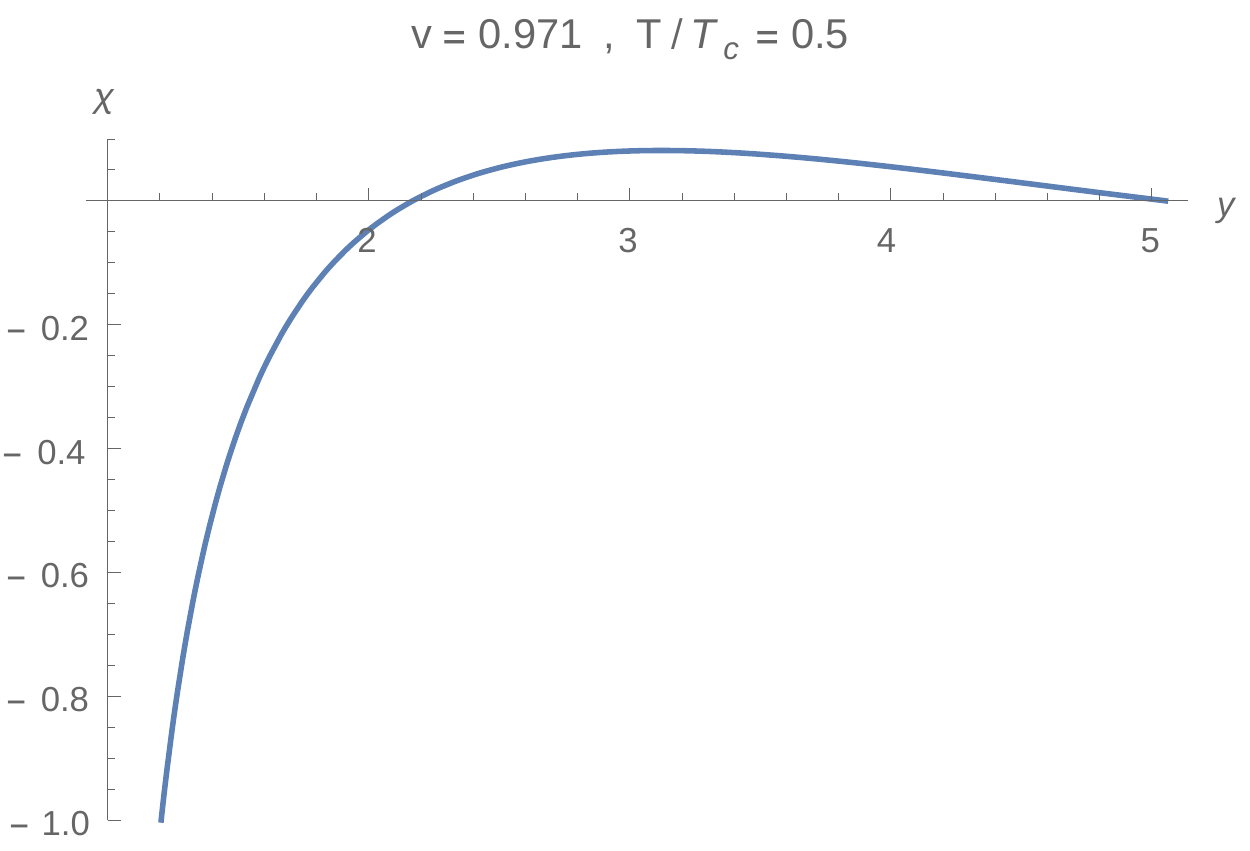}
\includegraphics[width=0.45\textwidth]{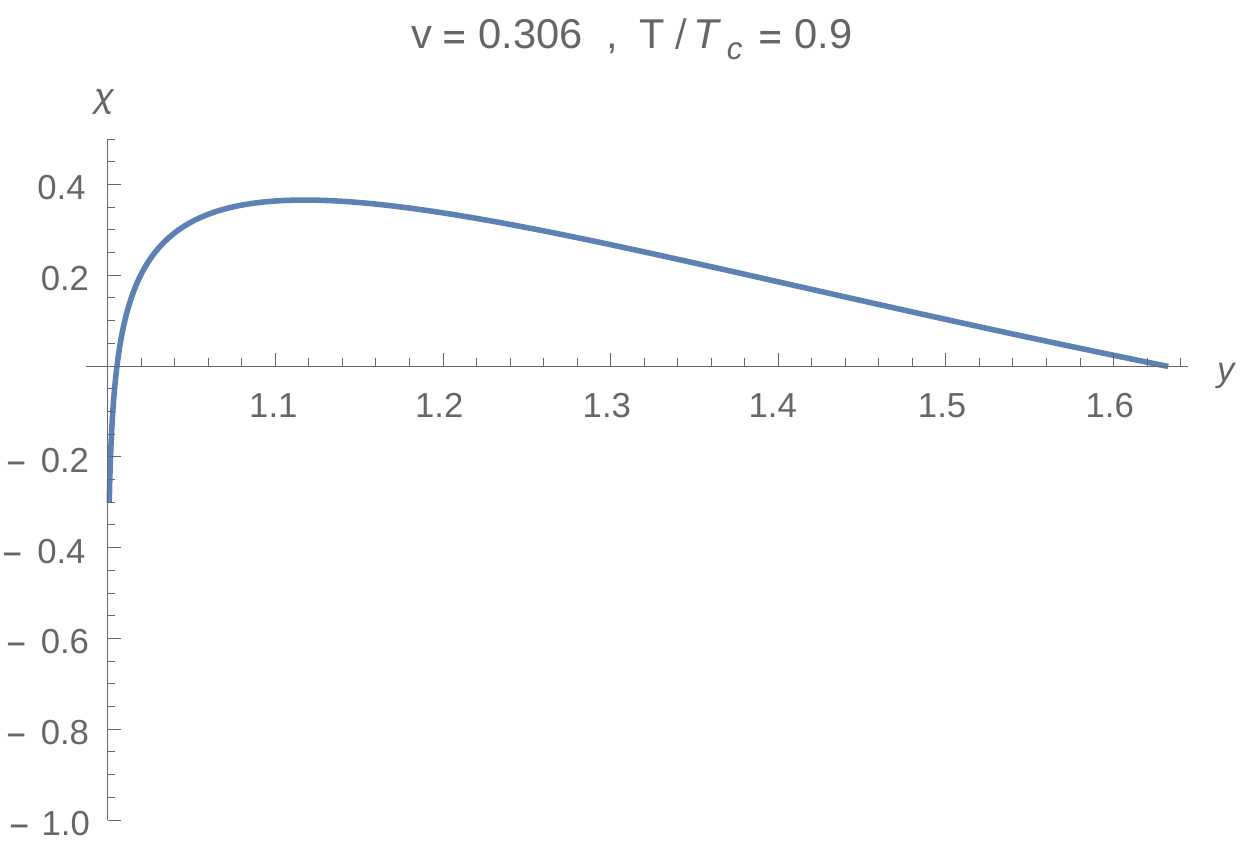}
\includegraphics[width=0.45\textwidth]{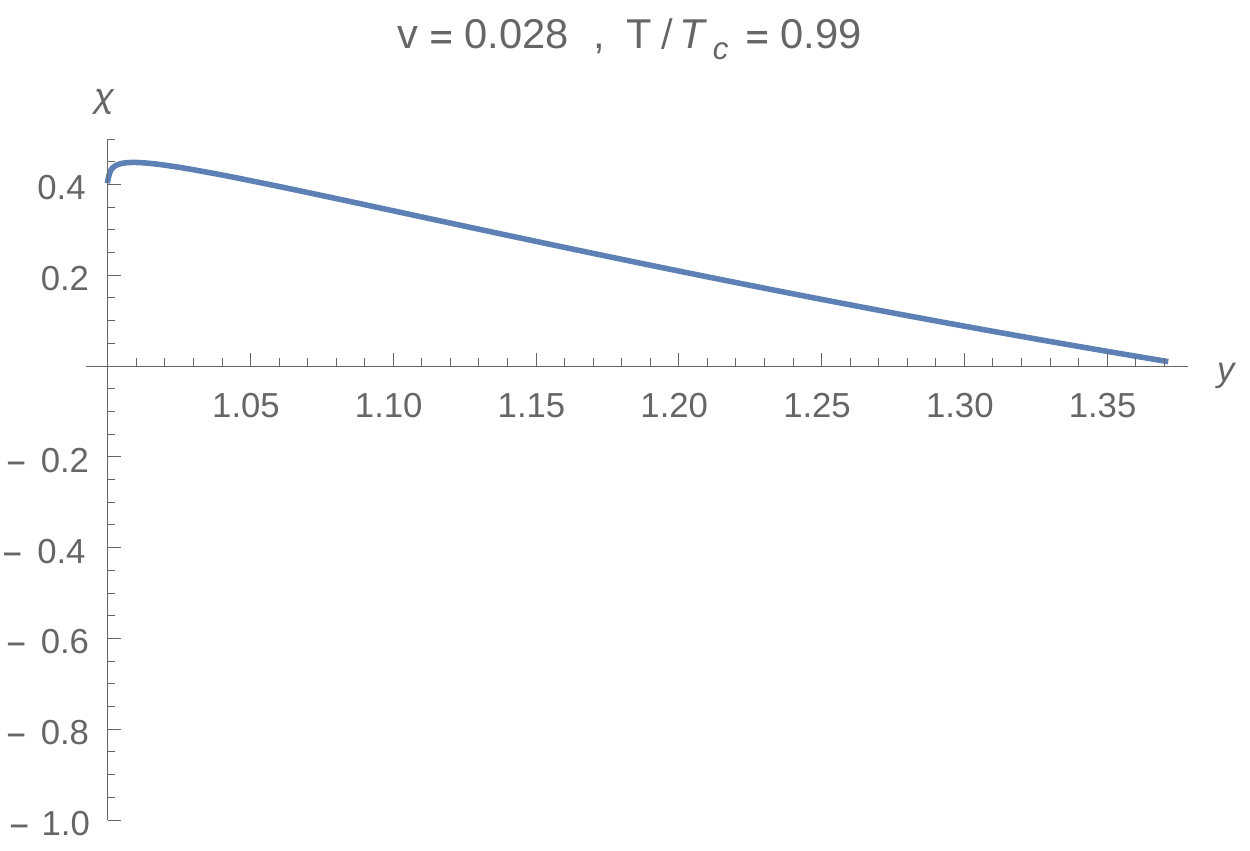}
\end{center}
\caption{Solutions for the wall profile $\chi \equiv \xi/L$ for different values of the velocity.}
\label{figsolnew}
\end{figure}

Plots of $\chi$ from (\ref{chiprimenew}) are reported in figure \ref{figsolnew}.
The solution resembles the trailing configuration of figure \ref{figprof}, with slide following the wall (which by definition is at $\chi=0$), for large velocity (when $u_* \sim u_J$), while for small velocity it resembles a solution with slide preceding the wall, up to $u_* \sim u_T$ where it turns around, in such a way that momentum flows towards the horizon.\footnote {Note that $y_J$ depends on $T/T_c$.}
Thus, at large velocity there is a sizeable part of the brane dual to the plasma (falling into the horizon) which is behind the wall, as pictured in figure \ref{trailing3d}.

\section{Steady-state bubbles in generic $Dp$-$Dq$-$\bar Dq$ setups}
\label{sec:general}

Chiral phase transitions in the deconfined phase of strongly coupled gauge theories with fundamental matter in different dimensions are quite a general feature in top-down holography. 
Here we focus on WSS-like setups involving $N_f$ $Dq$-$\bar Dq$ flavor branes probing the black hole background sourced by $N\gg1$ $Dp$-branes. Examples of these models both at zero and at finite temperature can be found in \cite{Kruczenski:2003uq,Rodriguez:2005jr,Gao:2006up,Antonyan:2006pg,Fujita:2016gmu,Gepner:2006qy}. 
These models exhibit chiral symmetry breaking in different dimensions and for different flavor content, even defect-like.
Thus, they allow probing formula (\ref{velocitygene}) in quite different conditions.

We are going to follow closely the steps made for the WSS model in sections \ref{sec:WSS}, \ref{sec:WSSsteady}. 
After introducing the backgrounds dual to the gluonic part of the models and the equation for the flavor bubble  steady state in section \ref{secbackgrounds}, we calculate the drag force, the total friction force and the bubble wall velocity in sections \ref{generalDpDqdrag}, \ref{generalDpDqvel}. 
In all the cases the result can be written as in formula (\ref{velocitygene}).

\subsection{Backgrounds and steady-state equation}
\label{secbackgrounds}
The black hole background sourced by the $Dp$-branes in string frame includes a metric and dilaton, which in the near-horizon limit read
\bea
&& ds^2 = H^{-1/2}\left[ -f dt^2+ dx_i dx_i\right] + H^{1/2}\left[\frac{du^2}{f} + u^2d\Omega_{8-p}^2\right]\,,\nonumber\\
&& H(u) = \left(\frac{R}{u}\right)^{7-p}\,,\quad f(u)=1-\left(\frac{u_T}{u}\right)^{7-p}\,,\quad e^{\Phi}=g_s H^{(3-p)/4}\,,
\label{back}
\eea
where $i=1,\dots , p$, $d\Omega_{8-p}^2$ is the metric of a unit $(8-p)$-sphere $S^{8-p}$ and $R$ is the background radius
\be
R^{7-p} = g_s N (4 \pi l_s^2)^{\frac{7-p}{2}} \frac{\Gamma(\frac{7-p}{2})}{4 \pi}\,.
\ee
The background also includes a Ramond-Ramond potential $C_{01\dots p}=H^{-1}$. 
The black hole temperature is given by
\be
T = \frac{7-p}{4\pi R}\left(\frac{u_T}{R}\right)^{(5-p)/2}\,.
\label{T}
\ee
One of the spacelike Minkowski coordinates, say $x_p$, can be chosen to be compact (like $x_4$ in WSS), but the analysis holds also in the non-compact case. Let us consider $Dq$-$\bar Dq$ flavor brane probes orthogonal to $x_p$ and placed at a certain distance $L$ from each other along that direction. Let us also assume that the flavor branes extend in $d$ spacelike Minkowski directions and along the radial direction $u$ and that they wrap an $n$-cycle in $S^{8-p}$, so that $q=d+n+1$. This introduces fundamental matter fields propagating on a $(d+1)$-dimensional defect in the dual field theory. 

These setups enjoy chiral symmetry breaking/preserving transitions at finite temperature with the same features as in the WSS model. Depending on the temperature being smaller or larger than some critical temperature $T_c\sim L^{-1}$, connected U-shaped flavor branes or disconnected stacks will provide the lowest energy configurations. We will consider bubbles of connected vacuum nucleated in the false disconnected vacuum.

In the thin-wall case, we can imagine a bounce solution analogous to the one found in WSS. We can thus repeat the same analysis as before describing the bubble wall through an ansatz of the form
\be
z = v t +\xi(u, \gamma)\,,\quad z\equiv x_d\,, \quad \gamma\equiv x_p\,.
\label{genwal}
\ee

Using the ansatz (\ref{genwal}) it is easy to realize that the DBI action for the $Dq$-brane profile describing the bubble wall 
reduces to
\be \label{genSw}
S_w = - \frac{k}{L} \int dt \, du \, d\gamma \, \left ( \frac{u}{R} \right )^{\frac{7-p}{4} (3+d-p-n)} u^{n} \sqrt{1 +  (\partial_\gamma \xi)^2 + \left ( \frac{u}{R} \right )^{7-p} f \, (\partial_u \xi)^2 - f^{-1} v^2}\,,
\ee
where
\be
k=\frac{T_q}{g_s} A L V(S^{n})\,,\quad A=\int dx^1 ... dx^{d-1}\,,\quad L\equiv \int_{worldvolume} d\gamma\,,
\ee
$T_q=(2\pi)^{-q} l_s^{-q-1}$ being the brane tension and $V(S^{n})=2\pi^{\frac{n+1}{2}}/\Gamma \pr{\frac{n+1}{2}}$ the volume of the $n$-cycle.

The Euler-Lagrange equations are
\be \label{genEL}
\partial_u \pi_\xi^u+ \partial_\gamma \pi_\xi^\gamma = 0\,,
\ee
for
\begin{subequations}
\bea \label{genpiu}
\pi_\xi^u =  k \frac{u^{n} f \left ( \frac{u}{R} \right )^{\frac{7-p}{4}[7+d-p-n]} \partial_u \xi}{ \sqrt{1 +  (\partial_\gamma \xi)^2 + \left ( \frac{u}{R} \right )^{7-p} f \, (\partial_u \xi)^2 - f^{-1} v^2}} \, , \\
\label{genpi4}
\pi_\xi^\gamma =  k \frac{u^{n} \left ( \frac{u}{R} \right )^{\frac{7-p}{4}[3+d-p-n]} 
\partial_\gamma \xi}{ \sqrt{1 +  (\partial_\gamma \xi)^2 + \left ( \frac{u}{R} \right )^{7-p} f \, (\partial_u \xi)^2 - f^{-1} v^2}}\, .
\ea
\end{subequations}
From the above equations it follows that
{\small
\be \label{genpartialuxi}
\partial_u \xi = \pi_\xi^u \left ( \frac{u}{R} \right )^{\frac{7-p}{4}[n+p-d-7]} f^{-\frac{1}{2}} \sqrt{\frac{1-f^{-1}v^2}{k^2 f u^{2n}-(\pi_\xi^u)^2 \left ( \frac{u}{R} \right )^{\frac{7-p}{2}[n+p-d-5]}- f \, (\pi_\xi^\gamma)^2 \left ( \frac{u}{R} \right )^{\frac{7-p}{2}[n+p-d-3]}}} \,
\ee
}
and
{\small
\be \label{genpartialgammaxi}
\partial_\gamma \xi = \pi_\xi^\gamma \left ( \frac{u}{R} \right )^{\frac{7-p}{4}[n+p-d-3]} f^{\frac{1}{2}} \sqrt{\frac{1-f^{-1}v^2}{k^2 f u^{2n}-(\pi_\xi^u)^2 \left ( \frac{u}{R} \right )^{\frac{7-p}{2}[n+p-d-5]}- f \, (\pi_\xi^\gamma)^2 \left ( \frac{u}{R} \right )^{\frac{7-p}{2}[n+p-d-3]}}} \, .
\ee
}

\subsection{The drag force}
\label{generalDpDqdrag}
Let us first notice that in (\ref{genpartialuxi}) and (\ref{genpartialgammaxi}) the numerator under square root vanishes when $f(u_c)=v^2$ i.e.~at
\be
u_c(v) = \frac{u_T}{(1-v^2)^{\frac{1}{7-p}}}.
\label{genuc}
\ee
Recalling the relation (\ref{T}) between $u_T$ and $T$, the above relation allows us to define a boosted temperature $T_{boost}$ related to $u_c$ and given by
\be
T_{boost}(v) = \frac{T}{(1-v^2)^{\frac{5-p}{2(7-p)}}} \, .
\label{teffgen}
\ee
Assuming the wall (excluded its boundaries) to have a trivial profile in $\gamma$ (the ``rectangular'' configuration), the value of the momentum $\pi_{\xi}^u$ in $u_c$, which we relate to the drag force as in the WSS case, will be thus given by 
\be
\pi_{\xi}^u(u_c) = k u_c^n v \left(\frac{u_c}{R}\right)^{\frac{7-p}{4}(5+d-p-n)}\,.
\ee
The drag force per unit area will read
\be
\frac{F_d}{A}=\frac{T_q}{g_s} V(S^n) L R^n \left(\frac{u_c}{R}\right)^{\frac{7-p}{4}(5+d-p-n)+n}v\,.
\ee
In order to rewrite the above expression in terms of field theory quantities let us first compute the free energy of the false vacuum. This is done by evaluating the renormalized on-shell DBI Euclidean action for the disconnected configuration
\be
S_E= 2\frac{T_q}{g_s}\frac{V(S^n)}{T}\int d^d x\int_{u_T}^{\infty} du u^n \left(\frac{u}{R}\right)^{\frac{7-p}{4}(d-n-p+3)}\,.
\label{discogen}
\ee
Removing the divergence at $u\rightarrow\infty$, we obtain the free energy density (hence minus the pressure) of the false vacuum in the dual field theory,
\be
f_f (T)= -p_f (T)= -2\frac{T_q}{g_s} V(S^n) R^{-\frac{7-p}{4}(d-n-p+3)}\frac{u_T^{\frac{7-p}{4}(d-n-p+3)+n+1}}{\frac{7-p}{4}(d-n-p+3)+n+1}\,.
\ee
In order to obtain the same observables at the boosted temperature $T_{boost}$, it suffices to replace $u_T$ with $u_c$ in the above expression. 

Exploiting the relation between $u_T$ and $T$ it is now easy to compute the enthalpy density $w=T\partial_T p$ of the false vacuum
\be
w_f = \frac{4}{5-p}\frac{T_q}{g_s} V(S^n) R^{-\frac{7-p}{4}(d-n-p+3)} u_T^{\frac{7-p}{4}(d-n-p+3)+n+1}\,.
\ee
 Putting all the above ingredients together we arrive at the following general expression for the drag force 
\be
\frac{F_d}{A} = \frac{\pi(5-p)}{(7-p)} L\, T_{boost}\, w_{false}(T_{boost})\, v\,,
\label{fdragrewgen}
\ee
which precisely reduces to (\ref{fdragexpr}) in the $p=4$ case. Let us notice that the overall coefficient
\be
b\equiv\frac{(5-p)}{(7-p)}\,,
\ee
corresponds to the power of $\gamma=(1-v^2)^{-1/2}$ in the relation (\ref{teffgen}) between $T_{boost}$ and $T$. This coefficient
carries information on the background gluonic radiation plasma probed by the flavor brane bubble. In particular we can express it as
\be
b = 2 \frac{p_{glue}}{w_{glue}} = 2 \frac{c_{s,glue}^2}{(1+c_{s,glue}^2)}\,,
\ee
where $p_{glue}, w_{glue}, c_{s,glue}$ are the pressure, enthalpy and speed of sound of the gluonic part of the plasma, respectively. In the WSS case they can be easily determined from eq.~(\ref{rhoradglue}). We thus see that the drag coefficient $C_d$ defined in (\ref{introfa}) can be written as
\be\label{cdekappa}
C_d = 2\pi \frac{p_{glue}}{w_{glue}} \kappa_{c} =2\pi \frac{c_{s,glue}^2}{(1+c_{s,glue}^2)}\kappa_{c} \,,
\ee
where
\be\label{kappaci}
\kappa_{c} \equiv L T_c\,,
\ee
is a model-dependent numerical coefficient relating $L$ to the critical temperature. 
The value of $\kappa_c$ is typically between $0.15$ and $0.3$ \cite{Gepner:2006qy}. 

\subsection{The bubble wall velocity}
\label{generalDpDqvel}
Adapting the analysis of section \ref{app:subb} to the general case, the zero-force condition in the steady state turns out to read
\be
F \equiv \frac{1}{L} \int d\gamma \, \pi_\xi^u (u_T, \gamma) = \overline{\pi_\xi^u (u_T)}=A\, \Delta p \,.
\ee
Details of the derivation of this formula, using the approximate setup with a rectangular connected configuration, are reported in appendix \ref{app:comgen}. With the same assumptions, we can extend the analysis of section \ref{subsec:wsszeroforce} to generic WSS-like models. In the rectangular approximation, $\pi^{\gamma}_{\xi}=0$, we can trade the boundary conditions (\ref{gammacond}) on $\pi_{\xi}^{\gamma}$ for source terms in the Euler-Lagrange equation. We thus get the equation
\be
-\partial_u \pi_\xi^u = 2 \frac{k}{L} u^{n} \left ( \frac{u}{R} \right )^{\frac{7-p}{4}[3+d-p-n]} \,,
\label{tobeinte}
\ee
where on the right-hand side one can recognize the Lagrangian density ${\cal L}_d$ of the disconnected configuration (\ref{discogen}). 
Therefore, it follows that
\be \label{genpartialpiu}
\pi_\xi^u = -  2 \, \frac{k}{L} \, R^{-(m-n-1)} \,  \frac{(u^m-u_{\ast}^m)}{m} \, ,
\ee
where we have fixed the integration constant requiring that $\pi_{\xi}^u(u_{\ast})=0$ and we have defined
\be
m \equiv \frac{7-p}{4} [ 3 + d - p - n ] + n + 1\,.
\label{defim}
\ee
Now, inserting (\ref{genpartialpiu}) in (\ref{genpartialuxi}) we obtain
\be \label{genxiprimedef}
\begin{split} 
\xi' &= - 2 \frac{k}{L} R^{-(m-n-1)} \frac{1}{m} \left ( u^m - u_{\ast}^m \right ) \left ( \frac{u}{R} \right )^{\frac{7-p}{4}(n+p-d-7)} f^{-1/2} \times \\
 &\times \sqrt{ \frac{1-f^{-1}v^2}{k^2 f u^{2n} - 4 \frac{k^2}{L^2} R^{-2(m-n-1)} \frac{1}{m^2} \left ( u^m - u_{\ast}^m \right )^2 \left ( \frac{u}{R} \right )^{\frac{7-p}{2}(n+p-d-5)} }} \, ,
\end{split}
\ee
as the generalization of  (\ref{xiprimenew}). Note again that $u_\ast \ge u_c$.

As in the WSS case, let us now consider a oversimplified thinnest-wall setup where 
the connected configuration, rigidly translating from the bubble center to the wall, is approximated by a rectangular one. Thus the trailing wall will have a horizontal line at $u=u_J$ trivially extending along $x_4$ as top boundary. Here we can assume
that $\partial_u \xi \rightarrow - \infty$, as $u \rightarrow u_J$ from above, so that, from (\ref{genpiu}), we get the boundary condition
\be
\pi_\xi^u(u_J) = - k u_J^{n} \sqrt{f(u_J)} \left ( \frac{u_J}{R} \right )^{\frac{7-p}{4}[5+d-p-n]}\equiv - k \mathcal{L}_{ch}(u_J) \,,
\ee
where we have recognized that the term in the middle corresponds to the Lagrangian density of the horizontal part of the connected configuration at $u=u_J$. 

Integrating eq.~(\ref{tobeinte}) between $u_c$ and $u_J$ we get, again under the assumption that the trailing wall has trivial profile along $\gamma$ away from its boundaries,
\be\label{deltapgen}
F_d\equiv \pi_\xi^u (u_c)= - k \, \mathcal{L}_{ch}(u_J) + 2 \frac{k}{L} \int_{u_c}^{u_J} du \, \mathcal{L}_d =  A \left(p_t(T)-p_f(T_{boost})\right)\,,
\ee
which is the expected generalized zero-force condition written in the form of eq. (\ref{fawssf}). As for the WSS case, we expect that the oversimplified setup can be extrapolated to the actual thinnest-wall configuration with a U-shaped connected solution describing the true vacuum. In this case, the pressure gradient written above will correspond to the realistic one. 
A complementary analysis of the above results can be found in appendix \ref{app:comgen}.
Together with (\ref{fdragrewgen}), formula (\ref{deltapgen}) reproduces equation (\ref{velocitygene}).

\section{Conclusions and discussion}
\label{sec:compare}

In this paper we have studied bubbles produced in  first-order chiral symmetry breaking transitions in strongly coupled quantum field theories with a top-down holographic description, in different dimensions.
We have focused on the late-time, steady-state configuration of a bubble of true vacuum expanding in the false vacuum plasma.
Modeling the steady state with a simplified configuration, we have derived formula (\ref{introfa}) for the total friction force and formula (\ref{introfapfp}) relating the pressure difference inside and outside the bubble to the friction exerted by the plasma on the bubble wall. Hence we have derived formula (\ref{velocitygene}) determining the bubble wall velocity. This formula is valid in all the considered cases, in every dimension, even for defect theories.
It has a very general form, the model dependence being encoded in a order-one numeric coefficient ($\kappa_c$ in (\ref{cdekappa}), (\ref{kappaci})).
As such, it could have a much broader validity than the present context.

A natural extension of this work would consist in performing a full-fledged numerical analysis of the non-linear PDE derived in section \ref{sec:WSSsteady} describing the steady state, in order to fully probe the validity of formula (\ref{velocitygene}) beyond the simplified configuration.
It would also be a very interesting task to study the same problem in different phase transitions in the flavor sector of holographic theories, as e.g.~the ones in \cite{Mateos:2006nu,Kobayashi:2006sb,Mateos:2007vn,Mateos:2007vc}, and possibly in holographic confining phase transitions.

Let us conclude by briefly comparing formula (\ref{velocitygene}) with the known results in the literature about the bubble wall velocity.
Microscopic computations of the velocity are actually extremely rare and often perturbative.
The first such computations, based on the Boltzmann equation and the equation for a scalar field, dates back to the works of Moore and Prokopec \cite{Moore:1995ua,Moore:1995si} for the electroweak phase transition, giving a runaway behavior.
For the same system, in 2017 Bodeker and Moore realized that a NLO particle contribution to the friction exerted by the plasma on the bubble wall is linearly proportional to the Lorentz factor $\gamma$ at large velocities \cite{Bodeker:2017cim}, allowing for an estimate of a non-runaway wall velocity.
Recently, there has been an attempt to extend these results to all perturbative orders, with the main effect that the linear $\gamma$ dependence would become a quadratic dependence \cite{Hoeche:2020rsg}.
It is highly non-trivial to adapt these perturbative results to strongly coupled physics, but a very rough estimate can be given along the lines of \cite{vonHarling:2019gme}, with the velocity expressions in the two cases (linear and quadratic $\gamma$ dependence) being
\be\label{BM} 
v_{L} \sim \sqrt{1- \left( \frac{T}{\Lambda} \right)^6}\,, \qquad 
v_{Q} \sim \sqrt{1- \left( \frac{T}{\Lambda} \right)^3}\,, \qquad ({\rm for \ large}\ v)\,.
\ee
In these formulae $\Lambda$ is the mass gap of the theory.

In the last year, another interesting study has derived the $\gamma$ dependence of the bubble dynamics in a simple way from covariant conservation of the energy-momentum tensor of the bubble-plasma system, with the result
\be\label{prokopec}
\Delta p = (1 - \gamma^2) T \Delta s\,,
\ee 
where 
$s$ is the entropy density \cite{Mancha:2020fzw}.
This result is reminiscent of our formula (\ref{velocitygene}), e.g.~in the form (\ref{effesuahere}).
The main difference is due to the first term in the r.h.s.~of (\ref{effesuahere}) (the drag force), which is  linear in $v$ at small velocities; such a linear behavior is absent in (\ref{prokopec}).
Note that formula (\ref{prokopec}) assumes that the plasma can be modeled as a perfect fluid, while no such approximation is made in (\ref{velocitygene}).

\begin{figure}[t!]
\begin{center}
\includegraphics[width=0.6\textwidth]{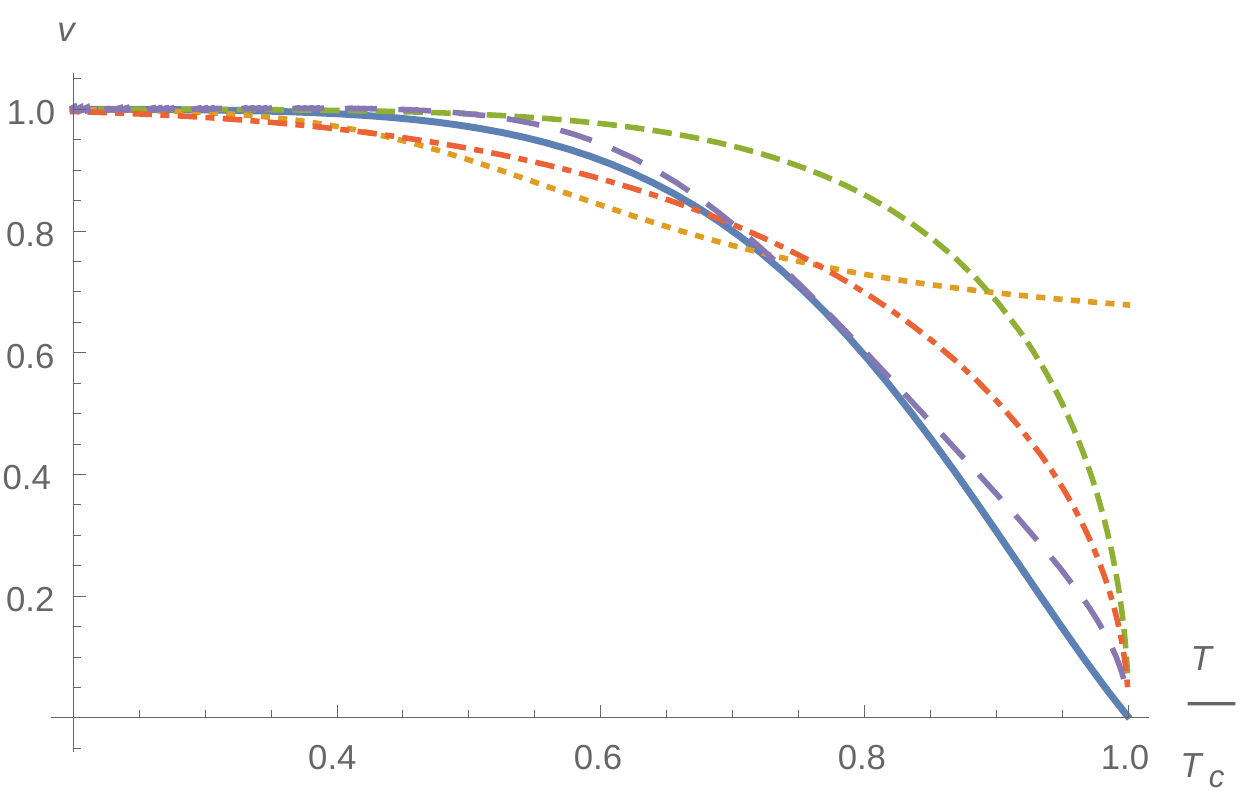}
\end{center}
\caption{Bubble velocity as a function of the relative temperature $T/T_c$ in the WSS model. The solid line is the result of this paper, formula (\ref{velocitygene}) with $C_d(WSS)=0.16$. The dotted line is the Chapman-Jouguet formula (\ref{CJ}) with parameters $\lambda=100, N_c=100, N_f=1, f_{\chi}/M_{KK}=10$, where $f_{\chi}$ is given in (\ref{fchil}) (this choice corresponds to a $v_{CJ}$ close to the other results for large $v$). The short-dashed (orange) and dot-dashed lines correspond to the linear and quadratic $\gamma$-dependences as in (\ref{BM}). The long-dashed (violet) curve corresponds to formula (\ref{prokopec}).}
\label{figvall}
\end{figure}

In the quantitative comparison below we also consider the Chapman-Jouguet formula, even if it does not correspond to a microscopic calculation, as a benchmark value very commonly employed in the calculation of gravitational wave spectra
\be\label{CJ}
v_{CJ} = \frac{1/\sqrt{3}+\sqrt{\alpha^2 + 2\alpha/3}}{1+\alpha}\ ,
\ee 
with
\be\label{alphacap}
\alpha = \frac{\Delta\theta}{\rho_{rad}}\ ,
\ee
where $\theta= (\rho-3p)/4$ is the trace of the energy-momentum tensor and $\rho_{rad}$ includes the energy density of the flavors as well as the one of the gluonic part of the plasma (\ref{rhoradglue}). In the WSS case in the probe approximation this parameter (to be eventually computed at the percolation temperature) is, in fact \cite{Bigazzi:2020avc},
\be
\alpha = \frac{\lambda^2}{480\pi^2}\frac{N_f}{N}\frac{T}{M_{KK}}\left (3+ T\partial_T\right) \Delta\tilde S\,.
\ee
In figure \ref{figvall} we compare quantitatively the various formulae for the velocity mentioned above in the specific case of the WSS model.
At large velocities the formulae give comparable results, while they differ significantly at intermediate and moderate velocities.
It is interesting to note that formula (\ref{prokopec}) gives an estimate that is quite close to our result even quantitatively, apart from the discrepancy in the small-velocity regime mentioned above.

Let us conclude by noticing that in the small-velocity limit, the total friction force (\ref{introfa}) to order $v^2$, reads
\be
\frac{F}{A}\approx \frac{2 p_{glue}}{w_{glue}} w_f(T) \left[\pi \kappa_c \frac{T}{T_c} v + \frac12 v^2\right]\,,
\label{F/Asmallv}
\ee
where we have used the general results of section \ref{sec:general}. The ${\cal O}(v)$ drag force contribution in the above expression, is complemented by a ${\cal O}(v^2)$ 
term which does not explicitely depend on the details of the transition and is proportional to the enthalpy density of the false vacuum. The structure of this term resembles the force contribution due to a ``snowplow'' effect \cite{Dine}, where a velocity-dependent density increase is induced just in front of the wall. The possibility that the total restraining force on the wall may result from both a purely dissipative contribution (the drag force) and ``local equilibrium" effects, related for instance to heating of the plasma in front of the wall, has been 
explored in the small-velocity regime in e.g. \cite{Konstandin:2010dm}.

Finally, notice that in the $v\ll1$ limit, to linear order in $v$, our formula (\ref{velocitygene}) gives
\be
v\approx C_d^{-1}\frac{T_c}{T}\frac{\Delta p}{w_f}\,.
\label{linearv}
\ee
The very recent investigation in a bottom-up holographic model in \cite{BCN} reports a possible linear relation (for $v\lesssim 0.3$) between the bubble velocity and the ratio of the pressure difference between the false and the true vacua, over the energy density of the false vacuum, in agreement with the above small-velocity result. The latter also agrees with standard expectations based on the hypothesis - which is actually realized in our setups according to eq. (\ref{F/Asmallv}) - that friction is dominated by a linear in $v$ effect at small velocities. 

\vskip 15pt \centerline{\bf Acknowledgments} \vskip 10pt 

\noindent 
We are deeply indebted to A. Paredes for comments and observations. We thank F.R. Ares, C.A. Cremonini, O. Henriksson, M. Hindmarsh, C. Hoyos, N. Jokela, D. Mateos, R. Rollo and G. Tallarita for comments and helpful discussions.


\appendix

\section{WSS zero-force condition: rectangular connected shape}
\label{explicitcheck2}

\begin{figure}[htb]
\begin{center}
\includegraphics[width=0.2\textwidth]{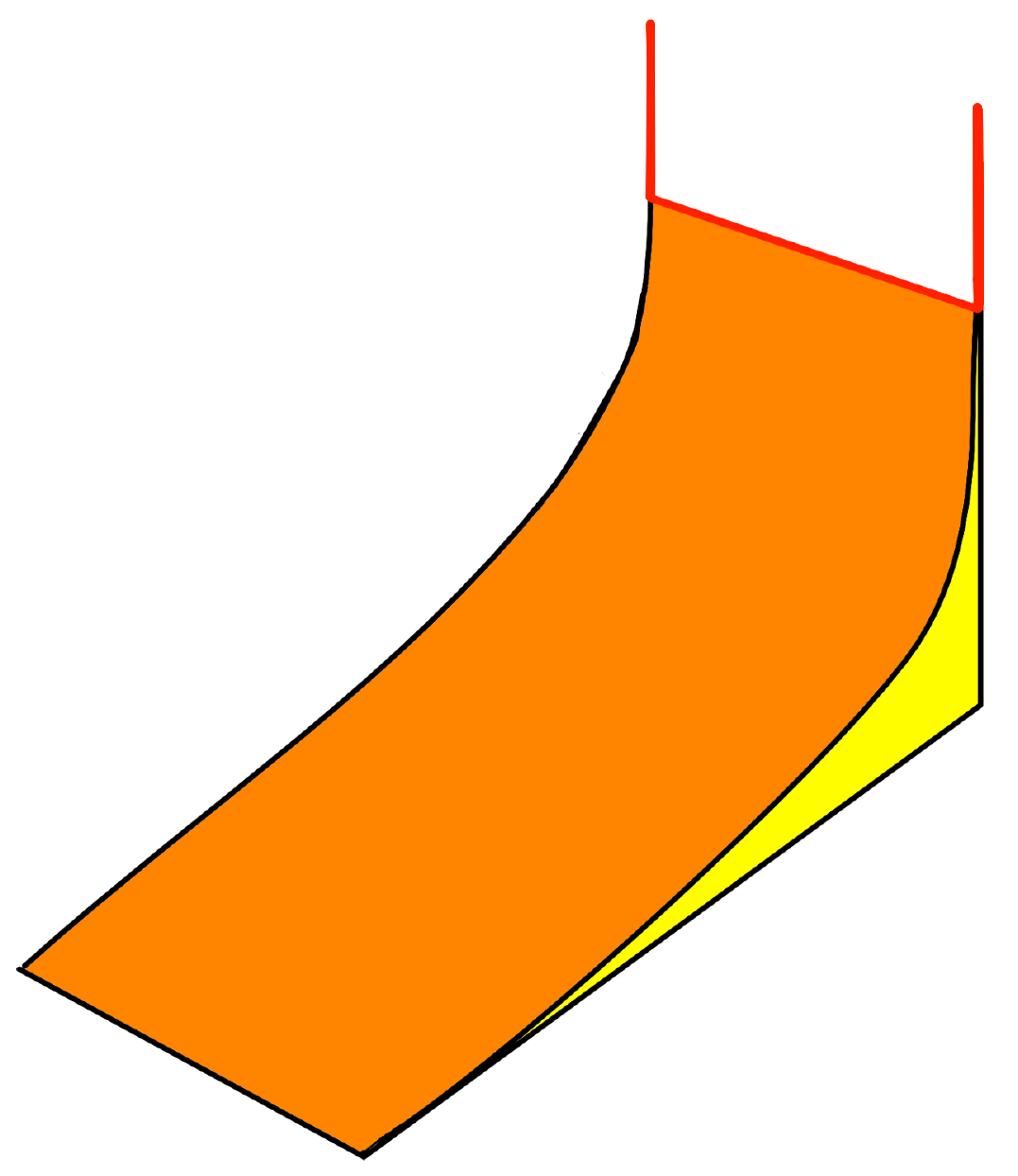}
\end{center}
\caption{A cartoon of the trailing wall and its boundaries in the thinnest-wall limit at $t\gg t_n$. In red the rectangular $D8$ whose horizontal part insists on the wall.}
\label{figwslides}
\end{figure}

Let us try to consider the complete $D8$-brane configuration, as represented in figure \ref{trailing3d} but for the fact that the smooth U-shaped piece is replaced by two vertical slides at constant $x_4 = \pm L/2$ and a horizontal one at $u=u_J$. 
We can split the total bounce action into six pieces associated to the six relevant brane branches: the connected solution on the left of the wall (which we split into the horizontal and the two vertical branches), the disconnected solution on the right (which, in turn, is conveniently split into the $u \geq u_J$ and the $u \leq u_J$ parts), the trailing wall itself (in orange in figure \ref{figwslides}) and the two vertical slides (in yellow in figure \ref{figwslides}) at $x_4=L/2$ and $x_4=-L/2$. The trailing profile is denoted by $z=z(u,x_4)$. The two vertical slides can be described as disconnected solutions extended from $z(u,\pm L/2)$ to $z(u_J,\pm L/2)$. 
Splitting the action as described above, we write it as
\be
S = S_{conn, h} + S_{conn,v} + S_{disconn, +} + S_{disconn, -} + S_w + S_{sl}  \ ,
\ee
where
\begin{subequations}
\ba
S_{conn,h} &=&  -\frac{k}{L} \int dt \int dx_4 \int_{u_T}^{+\infty} du\, \delta(u-u_J) \int_{-\infty}^{z_w(u_J,x_4)}dz \,{\cal L}_{c}\ , \\
S_{conn,v} &=&  -\frac{2 k}{L} \int dt \int dx_4 \, \delta(x_4-L/2) \int_{u_T}^{+\infty} du \, \Theta (u-u_J) \int_{-\infty}^{z_w(u_J,L/2)}dz \, \,{\cal L}_{d}\ , \q \\
S_{disconn, +} &=&  -\frac{2 k}{L} \int dt \int dx_4 \, \delta(x_4-L/2) \int_{u_T}^{+\infty} du \, \Theta (u-u_J) \int ^{\infty} _{z_w(u_J,L/2)} dz \, \,{\cal L}_{d}\ , \q\\
S_{sl}& =& - \frac{2k}{L}\int dt \int dx_4\, \delta(x_4-L/2) \int_{u_T}^{+ \infty} du \, \Theta (u_J-u) \int_{z_w(u,L/2)}^{z_w(u_J,L/2)} dz{\cal L}_d  \ , \q \\
\label{Sslides2}
S_{disconn, -} &=&  -\frac{2 k}{L} \int dt \int dx_4 \, \delta(x_4-L/2) \int_{u_T}^{+ \infty} du \, \Theta (u_J-u) \, \int ^{+\infty} _{z_w(u_J,L/2)} dz  \,{\cal L}_{d}\ ,\\
S_{w} &=& - \frac{k}{L}\int dt \int dx_4 \int_{u_T}^{+\infty}  du\, \Theta (u_J-u) \,{\cal L}_w \ ,
\ea
\end{subequations}
where
\be
k \equiv \frac{T_8}{g_s} A\,L\,V(S^4) \,, \quad A\equiv \int dx_1 dx_2\ .
\label{kdef2}
\ee
Here,
\be
\mc{L}_w =\left(\frac{u}{R}\right)^{-3/2}u^4 \sqrt{1 + (\partial_4\xi)^2+ f_T(u) \left(\frac{u}{R}\right)^3 (\partial_u \xi)^2 - f_T(u)^{-1}v^2}\ .
\label{sdbinewtotappb}
\ee

In the actions above, the overall factor $2$, when present, takes into account the two $x_4=\pm L/2$ branches of the embedding. In principle, we should write two different actions, one for each of the branches. We define the momenta as in (\ref{pixinew}) and (\ref{pi4new}).

It is important to stress that the domain of the embedding profile $z=z(u,x_4)$ is the region $D_{rect}$ spanned by $x_4 \in [-L/2,L/2]$ and $u \in [u_T,u_J]$.
Let us take the variation of this action with respect to $z_w (u)$, also varying the boundary values,
\ba
\label{variationtotalactionrectangular}
\d S &=&  \frac{1}{L} \int dt \int dx_4 \int_{u_T}^{+\infty}  du\, \Theta (u_J-u) (\p_u \pi_\xi ^u + \p_4 \pi_\xi ^4 )\, \d z_w (u) \nb \\
&-&\frac{1}{L}  \pr{k \int dt \int dx_4 \mc{L}_c + \int d t \int dx_4 \pi_\xi ^u}_{u=u_J} \d z_w (u_J,x_4) \nb \\
&+&\frac{1}{L}\int d t \int dx_4 \pi_\xi ^u |_{u=u_T} \d z_w (u_T) \nb  \\
&+&  \frac{1}{L} \int d t \int_{u_T} ^{u_J} du \pq{- \pi_\xi ^4 (x_4,u) + k \mc{L} _d}_{x_4=L/2} \d z_w (u,L/2) \nb \\
&+& \frac{1}{L} \int d t \int_{u_T} ^{u_J} du \pq{\pi_\xi ^4 (x_4,u) + k  \mc{L} _d}_{x_4=-L/2} \d z_w (u,-L/2)  \ ,
\ea
from which we read the Euler-Lagrange equation (valid for $u \leq u_J$)
\be
\label{ELquationalessioII}
\p_u \pi_\xi ^u + \p_4 \pi_\xi ^4 =0  \ .
\ee
Since we vary $S_{sl}$ and the trailing wall ends on the lateral slides, we have to take Neumann boundary conditions along $x_4$. Analogously for the boundary condition at $u=u_J$.
Thus, from the last two lines of (\ref{variationtotalactionrectangular}), we read
\be
\label{momentum4atboundary4}
\int_{u_T} ^{u_J} du \, \pi ^4 _\xi (x_4=L/2,u) = - \int_{u_T} ^{u_J} du \, \pi ^4 _\xi (x_4=-L/2,u) = k \int_{u_T} ^{u_J} du \,\mc{L} _d \ .
\ee
and from the second, 
\be
\label{momentumuatuj}
\frac{1}{L} \int dx_4 \pi_\xi ^u (x_4,u_J) = - k \mc{L}_c \ .
\ee
Integrating the Euler-Lagrange equation (\ref{ELquationalessioII}) over the region $D_{rect}$ and using (\ref{momentum4atboundary4}) and (\ref{momentumuatuj}) we find
\be
\label{g1appb}
\frac{1}{L}\int dx_4\, \pi_{\xi}^{u}(u_T, x_4) =  \frac{2k}{L}\int_{u_T}^{u_J}du{\cal L}_d - k {\cal L}_{c\,h}(u_J) \ ,
\ee
where we have used the fact that the horizontal red embedding does not depend on $x_4$. 
The left-hand side gives the friction force.
As we have shown in the main body, the right-hand side is proportional to the difference between the static rectangular connected and disconnected on-shell Euclidean actions and thus to the related pressure gradient. 
As a result, equation (\ref{g1appb}) is the zero-force condition
\be
\overline{\pi_{\xi}^u(u_T)} \equiv  \frac{1}{L}\int dx_4\, \pi_{\xi}^{u}(u_T, x_4)=  A \Delta p\ . 
\label{expe}
\ee

\section{The zero-force condition in the general case}
\label{app:comgen}
Let us consider the steady-state configuration for a generic WSS-like model. Let us also approximate the connected configuration on the left of the wall with a rectangular shape, having two vertical lines at $\gamma\pm L/2$ and a horizontal one at $u=u_J$. 
Under the same assumptions of section \ref{app:subb} it is natural to derive the zero-force condition in the following way. Let us be quite schematic so that we can avoid repeating the details given in section \ref{app:subb}. The contributions of the vertical slides of the trailing wall, where $\partial_{\gamma}\xi\rightarrow\pm\infty$ as $\gamma\rightarrow\pm L/2$ turn out to give the following boundary conditions (cfr.~also eq.~(\ref{genpi4}))
\be
\pi_\xi^\gamma\left(u, +\frac{L}{2}\right) = - \pi_\xi^\gamma\left(u, -\frac{L}{2}\right) = k \,  R^{-(m-n-1)} \, u^{m-1} \,,
\label{gammacond}
\ee
where the constant $m$ has been defined in (\ref{defim}). 
Moreover, the boundary condition at $u=u_J$, where $\partial_u \xi \rightarrow - \infty$ reads 
\be \label{piuuJ}
\pi_\xi^u (u_J) =  - k u_J^{n} \sqrt{f(u_J)} \left ( \frac{u_J}{R} \right )^{\frac{7-p}{4}[5+d-p-n]}\equiv - k \mathcal{L}_{ch}(u_J)   \, ,
\ee
where $\mathcal{L}_{ch}$ is proportional to the Lagrangian density of the horizontal part of the connected configuration.
Now, with analogous computations as those of section \ref{app:subb} we get
\be \label{genEL5}
\frac{1}{L} \int d\gamma \, \pi_\xi^u (u_T, \gamma) = - k \, \mathcal{L}_{ch}(u_J) + 2 \frac{k}{L} \int_{u_T}^{u_J} du \, \mathcal{L}_d \,,
\ee
which we read as the general zero-force condition where
\be
F \equiv \frac{1}{L} \int d\gamma \, \pi_\xi^u (u_T, \gamma) = \overline{\pi_\xi^u (u_T)} \, ,
\ee
and
\be \label{genDeltap1}
 k \, \mathcal{L}_{ch}(u_J) - 2 \frac{k}{L} \int_{u_T}^{u_J} du \, \mathcal{L}_d \equiv - A \Delta p\,.
\ee
Using known holographic maps between geometric data and QFT parameters we can rewrite the above condition, in the case with $N_f$ flavors, as 
{\small
\be
\label{finalgenDeltap}
\Delta p = J(p,n,d) \, \tilde L \, T^{\frac{2}{5-p}m} \, N N_f \, \left ( \frac{\lambda_{p+1}}{4 \pi} \right )^{\frac{n+p+d-5}{2(5-p)}} \left [ \frac{v}{(1-v^2)^{\frac{m}{7-p} + \frac{5-p}{2(7-p)}}} - \frac{2}{m} \frac{1}{\widetilde L} \left ( 1 - \frac{1}{(1-v^2)^{\frac{m}{7-p}}} \right )  \right ] \, ,
\ee}where
\be \label{genJWSS}
J(p,n,d) \equiv \frac{4 \pi^{\frac{n+3}{2}}}{\Gamma\left( \frac{n+1}{2} \right)} \left ( \frac{4 \pi}{7-p} \right )^{\frac{2}{5-p} m} \left [ \frac{\Gamma(\frac{7-p}{2})}{\pi^{\frac{7-p}{2}} 4 \pi} \right ]^{\frac{(n+5)+(d-p)}{2(5-p)}} \,,
\ee
$\lambda_{p+1}=4\pi g_s N (2\pi l_s)^{p-3}$ is the 't Hooft coupling of the $SU(N)$ theory and
\be \label{genLtilde}
\widetilde L = \frac{4 \pi}{7-p} \, L \, T \, .
\ee
This is the generalized version of eq. (\ref{effesuahere}).



\begin{thebibliography}{99}


\bibitem{Caprini:2019egz}
C.~Caprini, M.~Chala, G.~C.~Dorsch, M.~Hindmarsh, S.~J.~Huber, T.~Konstandin, J.~Kozaczuk, G.~Nardini, J.~M.~No and K.~Rummukainen, \textit{et al.}
``Detecting gravitational waves from cosmological phase transitions with LISA: an update,''
JCAP \textbf{03}, 024 (2020)
[arXiv:1910.13125 [astro-ph.CO]].

\bibitem{Hindmarsh:2020hop}
M.~B.~Hindmarsh, M.~L\"uben, J.~Lumma and M.~Pauly,
``Phase transitions in the early universe,''
SciPost Phys. Lect. Notes \textbf{24}, 1 (2021)
[arXiv:2008.09136 [astro-ph.CO]].

\bibitem{Bigazzi:2020phm}
F.~Bigazzi, A.~Caddeo, A.~L.~Cotrone and A.~Paredes,
``Fate of false vacua in holographic first-order phase transitions,''
JHEP \textbf{12}, 200 (2020)
[arXiv:2008.02579 [hep-th]].

\bibitem{Bigazzi:2020avc}
F.~Bigazzi, A.~Caddeo, A.~L.~Cotrone and A.~Paredes,
``Dark Holograms and Gravitational Waves,''
JHEP \textbf{04}, 094 (2021)
[arXiv:2011.08757 [hep-ph]].

\bibitem{creminelli}P.~Creminelli, A.~Nicolis and R.~Rattazzi,
``Holography and the electroweak phase transition,''
JHEP \textbf{03}, 051 (2002)
[arXiv:hep-th/0107141 [hep-th]].
  
\bibitem{Khlebnikov:1992bx}
S.~Y.~Khlebnikov,
``Fluctuation - dissipation formula for bubble wall velocity,''
Phys. Rev. D \textbf{46}, R3223-R3226 (1992).
  
\bibitem{Arnold:1993wc}
P.~B.~Arnold,
``One loop fluctuation - dissipation formula for bubble wall velocity,''
Phys. Rev. D \textbf{48}, 1539-1545 (1993)
[arXiv:hep-ph/9302258 [hep-ph]].

\bibitem{Dine}
M.~Dine, R.~G.~Leigh, P.~Y.~Huet, A.~D.~Linde and D.~A.~Linde,
``Towards the theory of the electroweak phase transition,''
Phys. Rev. D \textbf{46}, 550-571 (1992)
[arXiv:hep-ph/9203203 [hep-ph]].

\bibitem{Liu}
B.~H.~Liu, L.~D.~McLerran and N.~Turok,
``Bubble nucleation and growth at a baryon number producing electroweak phase transition,''
Phys. Rev. D \textbf{46} (1992), 2668-2688.

\bibitem{Moore:1995ua}
G.~D.~Moore and T.~Prokopec,
``Bubble wall velocity in a first order electroweak phase transition,''
Phys. Rev. Lett. \textbf{75}, 777-780 (1995)
[arXiv:hep-ph/9503296 [hep-ph]].

\bibitem{Moore:1995si}
G.~D.~Moore and T.~Prokopec,
``How fast can the wall move? A Study of the electroweak phase transition dynamics,''
Phys. Rev. D \textbf{52}, 7182-7204 (1995)
[arXiv:hep-ph/9506475 [hep-ph]].

\bibitem{Espinosa:2010hh}
J.~R.~Espinosa, T.~Konstandin, J.~M.~No and G.~Servant,
``Energy Budget of Cosmological First-order Phase Transitions,''
JCAP \textbf{06}, 028 (2010)
[arXiv:1004.4187 [hep-ph]].

\bibitem{Konstandin:2014zta}
T.~Konstandin, G.~Nardini and I.~Rues,
``From Boltzmann equations to steady wall velocities,''
JCAP \textbf{09}, 028 (2014)
[arXiv:1407.3132 [hep-ph]].

\bibitem{Bodeker:2017cim}
D.~Bodeker and G.~D.~Moore,
``Electroweak Bubble Wall Speed Limit,''
JCAP \textbf{05}, 025 (2017)
[arXiv:1703.08215 [hep-ph]].

\bibitem{Dorsch:2018pat}
G.~C.~Dorsch, S.~J.~Huber and T.~Konstandin,
``Bubble wall velocities in the Standard Model and beyond,''
JCAP \textbf{12}, 034 (2018)
[arXiv:1809.04907 [hep-ph]].

\bibitem{Mancha:2020fzw}
M.~Barroso Mancha, T.~Prokopec and B.~Swiezewska,
``Field-theoretic derivation of bubble-wall force,''
JHEP \textbf{01}, 070 (2021)
[arXiv:2005.10875 [hep-th]].

\bibitem{Hoeche:2020rsg}
S.~H\"oche, J.~Kozaczuk, A.~J.~Long, J.~Turner and Y.~Wang,
``Towards an all-orders calculation of the electroweak bubble wall velocity,''
JCAP \textbf{03}, 009 (2021)
[arXiv:2007.10343 [hep-ph]].

\bibitem{Vanvlasselaer:2020niz}
A.~Azatov and M.~Vanvlasselaer,
``Bubble wall velocity: heavy physics effects,''
JCAP \textbf{01}, 058 (2021)
[arXiv:2010.02590 [hep-ph]].

\bibitem{Cai:2020djd}
R.~G.~Cai and S.~J.~Wang,
``Effective picture of bubble expansion,''
JCAP \textbf{03}, 096 (2021)
[arXiv:2011.11451 [astro-ph.CO]].

\bibitem{Konstandin:2010dm}
T.~Konstandin and J.~M.~No,
``Hydrodynamic obstruction to bubble expansion,''
JCAP \textbf{02}, 008 (2011)
[arXiv:1011.3735 [hep-ph]].

\bibitem{Balaji:2020yrx}
S.~Balaji, M.~Spannowsky and C.~Tamarit,
``Cosmological bubble friction in local equilibrium,''
[arXiv:2010.08013 [hep-ph]].

\bibitem{Mateos:2006nu}
D.~Mateos, R.~C.~Myers and R.~M.~Thomson,
``Holographic phase transitions with fundamental matter,''
Phys. Rev. Lett. \textbf{97}, 091601 (2006)
[arXiv:hep-th/0605046 [hep-th]].

\bibitem{Kobayashi:2006sb}
S.~Kobayashi, D.~Mateos, S.~Matsuura, R.~C.~Myers and R.~M.~Thomson,
``Holographic phase transitions at finite baryon density,''
JHEP \textbf{02}, 016 (2007)
[arXiv:hep-th/0611099 [hep-th]].

\bibitem{Mateos:2007vn}
D.~Mateos, R.~C.~Myers and R.~M.~Thomson,
``Thermodynamics of the brane,''
JHEP \textbf{05}, 067 (2007)
[arXiv:hep-th/0701132 [hep-th]].

\bibitem{Mateos:2007vc}
D.~Mateos, S.~Matsuura, R.~C.~Myers and R.~M.~Thomson,
``Holographic phase transitions at finite chemical potential,''
JHEP \textbf{11}, 085 (2007)
[arXiv:0709.1225 [hep-th]].

\bibitem{Witten:1998zw}
E.~Witten,
``Anti-de Sitter space, thermal phase transition, and confinement in gauge theories,''
Adv. Theor. Math. Phys. \textbf{2}, 505-532 (1998)
[arXiv:hep-th/9803131 [hep-th]].

\bibitem{Sakai:2004cn}
T.~Sakai and S.~Sugimoto,
``Low energy hadron physics in holographic QCD,''
Prog. Theor. Phys. \textbf{113}, 843-882 (2005)
[arXiv:hep-th/0412141 [hep-th]].


\bibitem{BCN}
Y.~Bea, J.~Casalderrey-Solana, T.~Giannakopoulos, D.~Mateos, M.~Sanchez-Garitaonandia and M.~Zilh\~ao,
``Bubble Wall Velocity from Holography,''
[arXiv:2104.05708 [hep-th]].

\bibitem{Ares:2020lbt}
F.~R.~Ares, M.~Hindmarsh, C.~Hoyos and N.~Jokela,
``Gravitational waves from a holographic phase transition,''
JHEP \textbf{04}, 100 (2021)
[arXiv:2011.12878 [hep-th]].

\bibitem{Aharony:2006da}
O.~Aharony, J.~Sonnenschein and S.~Yankielowicz,
``A Holographic model of deconfinement and chiral symmetry restoration,''
Annals Phys. \textbf{322}, 1420-1443 (2007)
[arXiv:hep-th/0604161 [hep-th]].

\bibitem{Antonyan:2006vw}
E.~Antonyan, J.~A.~Harvey, S.~Jensen and D.~Kutasov,
``NJL and QCD from string theory,''
[arXiv:hep-th/0604017 [hep-th]].

\bibitem{Bigazzi:2014qsa}
F.~Bigazzi and A.~L.~Cotrone,
``Holographic QCD with Dynamical Flavors,''
JHEP \textbf{01}, 104 (2015)
[arXiv:1410.2443 [hep-th]].

\bibitem{Coleman:1977py}
S.~R.~Coleman,
``The Fate of the False Vacuum. 1. Semiclassical Theory,''
Phys. Rev. D \textbf{15}, 2929-2936 (1977).

\bibitem{Herzog:2006gh}
C.~P.~Herzog, A.~Karch, P.~Kovtun, C.~Kozcaz and L.~G.~Yaffe,
``Energy loss of a heavy quark moving through N=4 supersymmetric Yang-Mills plasma,''
JHEP \textbf{07}, 013 (2006)
[arXiv:hep-th/0605158 [hep-th]].

\bibitem{Gubser:2006bz}
S.~S.~Gubser,
``Drag force in AdS/CFT,''
Phys. Rev. D \textbf{74}, 126005 (2006)
[arXiv:hep-th/0605182 [hep-th]].

\bibitem{Janiszewski:2011ue}
S.~Janiszewski and A.~Karch,
``Moving Defects in AdS/CFT,''
JHEP \textbf{11}, 044 (2011)
[arXiv:1106.4010 [hep-th]].

\bibitem{FuiniJohnF:2011aa}
J.~F.~Fuini, III and A.~Karch,
``Energy Loss Calculations of Moving Defects for General Holographic Metrics,''
Phys. Rev. D \textbf{85}, 066006 (2012)
[arXiv:1112.2747 [hep-th]].


\bibitem{Kruczenski:2003uq}
M.~Kruczenski, D.~Mateos, R.~C.~Myers and D.~J.~Winters,
``Towards a holographic dual of large N(c) QCD,''
JHEP \textbf{05}, 041 (2004)
[arXiv:hep-th/0311270 [hep-th]].

\bibitem{Rodriguez:2005jr}
M.~J.~Rodriguez and P.~Talavera,
``A 1+1 field theory spectrum from M theory,''
[arXiv:hep-th/0508058 [hep-th]].



\bibitem{Gao:2006up}
Y.~h.~Gao, W.~s.~Xu and D.~f.~Zeng,
``NGN, QCD(2) and chiral phase transition from string theory,''
JHEP \textbf{08}, 018 (2006)
[arXiv:hep-th/0605138 [hep-th]].



\bibitem{Antonyan:2006pg}
E.~Antonyan, J.~A.~Harvey and D.~Kutasov,
``Chiral symmetry breaking from intersecting D-branes,''
Nucl. Phys. B \textbf{784}, 1-21 (2007)
[arXiv:hep-th/0608177 [hep-th]].


\bibitem{Fujita:2016gmu}
M.~Fujita, C.~M.~Melby-Thompson, R.~Meyer and S.~Sugimoto,
``Holographic Chern-Simons Defects,''
JHEP \textbf{06}, 163 (2016)
[arXiv:1601.00525 [hep-th]].




\bibitem{Gepner:2006qy}
D.~Gepner and S.~S.~Pal,
``Chiral symmetry breaking and restoration from holography,''
[arXiv:hep-th/0608229 [hep-th]].

\bibitem{vonHarling:2019gme}
B.~Von Harling, A.~Pomarol, O.~Pujol\`as and F.~Rompineve,
``Peccei-Quinn Phase Transition at LIGO,''
JHEP \textbf{04}, 195 (2020)
[arXiv:1912.07587 [hep-ph]].

\end{thebibliography}
\end{document}